\documentclass[12pt]{article}
\usepackage{latexsym}
\usepackage{graphicx}
\def\dslash{\raisebox{1pt}{$\slash$} \hspace{-7pt} \partial}
\def\pslash{\raisebox{1pt}{$\slash$} \hspace{-7pt} p}

\def\Dslash{\hspace{3pt}\raisebox{1pt}{$\slash$} \hspace{-9pt} D}
\def\Dslashn{\hspace{3pt}\raisebox{1pt}{$\slash$} \hspace{-7pt} D}
\def\bea{\begin{eqnarray}}
\def\eea{\end{eqnarray}}
\def\be{\begin{equation}}
\def\ee{\end{equation}}
\def\nn{\nonumber}
\def\a{& \hspace{-7pt}}
\def\s{\hspace{1pt}}
\def\Z{{\bf Z}}
\begin{document}

\thispagestyle{empty}

\begin{center}
\hfill CERN-TH/2003-028\\
\hfill ROMA-1354/03 \\
\hfill SISSA-32/2003/EP \\

\begin{center}

\vspace{1.7cm}

{\LARGE\bf Electroweak symmetry breaking and \\[2mm]
fermion masses from extra dimensions}

\end{center}

\vspace{1.4cm}

{\bf Claudio A. Scrucca$^{a}$, Marco Serone$^{b}$ and
Luca Silvestrini$^{c}$}\\

\vspace{1.2cm}

${}^a\!\!$
{\em Theor. Phys. Div., CERN, CH-1211 Geneva 23, Switzerland}
\vspace{.3cm}

${}^b\!\!$
{\em ISAS-SISSA and INFN, Via Beirut 2-4, I-34013 Trieste, Italy}
\vspace{.3cm}

${}^c\!\!$
{\em INFN, Sez. di Roma, Dip. di Fisica, Univ. di Roma ``La Sapienza''} \\
{\em P.le Aldo Moro 2, I-00185, Rome, Italy}
\end{center}

\vspace{0.8cm}

\centerline{\bf Abstract}
\vspace{2 mm}
\begin{quote}\small

  We study higher-dimensional non-supersymmetric orbifold models where
  the Higgs field is identified with some internal component of a
  gauge field. We address two important and related issues that
  constitute severe obstacles towards model building within this type
  of constructions: the possibilities of achieving satisfactory Yukawa
  couplings and Higgs potentials. We consider models where matter
  fermions are localized at the orbifold fixed-points and couple to
  additional heavy fermions in the bulk. When integrated out, the
  latter induce tree-level non-local Yukawa interactions and a quantum
  contribution to the Higgs potential that we explicitly evaluate and
  analyse. The general features of these highly constrained models are
  illustrated through a minimal but potentially realistic
  five-dimensional example.  Finally, we discuss possible cures for
  the persisting difficulties in achieving acceptable top and Higgs
  masses.  In particular, we consider in some detail the effects
  induced in these models by adding localized kinetic terms for gauge
  fields.
\end{quote}

\vfill

\newpage

\setcounter{equation}{0}

\section{Introduction}

The Standard Model (SM) of electroweak and strong interactions is
extremely successful in reproducing all the available experimental
data up to currently accessible energies. However, the problem of
stabilizing the electroweak scale against quadratically divergent
radiative corrections to the Higgs mass suggests the
presence of new physics at scales not much larger than the $Z$ mass.
While supersymmetric extensions of the SM technically
solve the hierarchy problem and suggest the unification of gauge
couplings, the ever-rising lower bounds on supersymmetric particles
have recently stimulated studies of alternative methods of supersymmetry
and electroweak breaking based on the presence of extra dimensions
at the TeV scale \cite{Antoniadis}.

Higher-dimensional models open up the interesting possibility that the
Higgs boson arises from the internal component of a gauge field, with
a dynamics protected by higher-dimensional gauge invariance and
controlled in a predictive way by a higher-dimensional effective
theory. Such a possibility, already advocated several years ago
\cite{early}, has recently received renewed interest
\cite{Hatanaka:1998yp}--\cite{Burdman:2002se}. In
the case of toroidal internal spaces, the effective potential turns
out to be free of quadratic divergences, and even in
phenomenologically more viable orbifold models \cite{Dixon}, power
divergences arising through operators localized at the fixed-points
are strongly constrained \cite{vonGersdorff}. Exploiting these facts,
one can construct interesting orbifold models in which electroweak
interactions are unified in a larger group, which is broken through the
orbifold projection. Gauge bosons and Higgs fields arise respectively
from the four-dimensional and internal components of the
higher-dimensional gauge bosons. Matter fields can be introduced
either as bulk fields in representations of the unified group $G$, or
as boundary fields localized at the fixed-points where this is broken
to a subgroup $H$. The construction of realistic models of this type
is, however, a difficult task, because Higgs interactions are
severely constrained by the higher-dimensional gauge invariance.
In particular, achieving satisfactory flavour structures and electroweak
symmetry breaking at the same time represents the main problem in
this class of theories.

Flavour symmetry breaking can be achieved essentially in two ways,
depending on whether standard matter fields are introduced in the bulk
or at the fixed-points of the orbifold, i.e. the boundaries.  In
the case of bulk matter fields, standard Yukawa couplings can
originate only from higher-dimensional gauge couplings, which are
flavour-symmetric and completely determined by the group
representation. A possible way of obtaining couplings with a non-trivial
flavour structure consists in switching on mass terms with odd
coefficients and introducing mixings between the bulk matter fields
and additional boundary fields in such a way that unwanted light
fields eventually decouple \cite{Burdman:2002se}.  In the case of
boundary matter fields, standard Yukawa couplings cannot be directly
introduced, because they would violate the higher-dimensional symmetry
that protects the potential and therefore induce quadratic divergences,
just as in the SM \cite{vonGersdorff}. The only invariant interactions
that can be used are non-local interactions involving Wilson lines
\cite{Hall:2001zb}; an interesting possibility to generate this kind
of interactions was suggested in \cite{Csaki:2002ur} and consists
in introducing mixings between the matter fields and additional heavy
fermions in the bulk, which are then integrated out. These two flavour
symmetry breaking mechanisms are very similar from the microscopic
point of view, and rely basically on a mixing of bulk and boundary
fields. Interestingly, they also provide a natural explanation for the
large hierarchy of fermion masses, since they are exponentially
sensitive on some parameters of the microscopic theory.

Electroweak symmetry breaking occurs radiatively in this class of
models, and is equivalent to a Wilson line symmetry breaking
\cite{Wilson-line,hos} (or Scherk--Schwarz twist \cite{SS}), which
reduces the rank of $H$ (as required for electroweak symmetry
breaking) if the Wilson line and the embedding of the orbifold twist
do not commute. Since the tree-level Higgs potential is strongly
constrained by gauge invariance, any theory of this type is
potentially very predictive and constrained, even though it is only
an effective theory. The generic predictive power of the theory
depends on the parameter $M_c/\Lambda$, which controls the effect
of non-renormalizable operators at the compactification scale $M_c$.
$\Lambda$ is the physical cut-off scale, whose actual value can be
roughly estimated by using Na\"{\i}ve Dimensional Analysis (NDA)
\cite{NDA1,NDA2}. There are, however, special quantities, such as
the ratio of Higgs and gauge boson masses, that turn out to be quite
insensitive to $\Lambda$, because of the non-local nature of the
electroweak symmetry breaking in the internal dimensions.

The construction of phenomenologically viable models requires a
detailed and combined analysis of the flavour and electroweak symmetry
breaking mechanisms, since the two are closely connected.
Such an analysis has never been performed so far, and the aim of this
work is to fill this gap with an investigation of the possibility of
achieving realistic effective Yukawa couplings and at the same time a
satisfactory Higgs potential.  We focus for simplicity on five-dimensional
(5D) models,
where the tree-level Higgs potential is actually vanishing and
electroweak symmetry breaking is thus entirely governed, at
sufficiently weak coupling, by the one-loop Higgs effective potential.
The contribution to the latter arising from gauge fields or bulk
fermions in fundamental or adjoint representations is easily computed
and well-known \cite{hos,Kubo:2001zc,Haba:2002py}. In the presence
of a flavour symmetry breaking mechanism of the types described above,
however, the computation of the matter contribution to the effective
potential is more involved, since in both cases there are simultaneously
bulk fermions and boundary fermions mixing among each other. Putting
matter fields in the bulk as suggested in \cite{Burdman:2002se}, the
situation is further worsened by the need of odd mass terms that
distort the wave functions of the bulk fields on their own. For this
reason, we consider here the case suggested in \cite{Csaki:2002ur},
with matter fields located at the fixed-points.

The basic construction that we study consists of a pair of
left-handed and right-handed matter fermions located at possibly
different fixed-points and a heavy bulk fermion with quantum numbers
allowing couplings to both boundary fermions, in such a way that
non-local Wilson line effective interactions between the two can be
generated. We explicitly compute the tree-level effective action that
is induced for the matter fermions by integrating out the massive
bulk fermions. The result consists of a mass term, a wave-function
correction, and an infinite series of higher-derivative interactions.
All these terms are directly proportional to the bulk-to-boundary
couplings and exponentially sensitive to the bulk masses. At energies
much below the bulk mass, higher-derivative terms can be neglected and
after rescaling the wave-function one finds a physical mass that is
bounded in size. At energies much above the bulk mass, on the
contrary, the induced interactions get exponentially suppressed
by their derivative dependence. We also analyse the full one-loop
effective potential for the Higgs field induced by charged bulk
fields. Its dependence on the bulk-to-boundary couplings can be
reinterpreted in terms of diagrams involving boundary fields and the
momentum-dependent effective vertex between Higgs and boundary fields
that is induced at tree level by the bulk fermion. The soft behaviour
of this effective vertex at high momenta ensures that the loop integral
is free of any divergence. The full one-loop effective potential for
the Higgs fields is therefore finite, as expected on symmetry
grounds\footnote{This should be contrasted to \cite{Csaki:2002ur},
  where a Wilson line Yukawa coupling was introduced as a fundamental
  coupling in the tree-level Lagrangian and found to induce
  divergences in the one-loop Higgs potential.  The distinction
  between our situation and that of \cite{Csaki:2002ur} lies in the
  locality of the fundamental Lagrangian. If the latter is non-local,
  then quantum divergences are allowed to appear even in non-local
  operators (such as the Higgs mass in these models), since all of them
  must now be introduced in the theory from the beginning as
  independent couplings and can be used as counterterms.}. Moreover,
it turns out to have non-trivial symmetry-breaking minima. This shows
that the flavour and electroweak symmetry breaking mechanisms that we
consider are indeed compatible.

The above building blocks can be easily used to construct models that
incorporate all SM fermions and Yukawa couplings and offer therefore a
realistic arena for more quantitative investigations.  The only
universal quantity to play with is the radius $R$, which can be
considered as a parameter. In phenomenologically viable models, it must
be small enough to satisfy the current experimental bounds (see e.g.
\cite{Delgado:1999sv} and references therein), and the Vacuum
Expectation Value (VEV) $\alpha$ of the Wilson line phase must
therefore be small as well in order to match the mass of the $W$ bosons,
given by $m_W=\alpha/R$.  For the minimal version of the model, with
the simplest possible choice of gauge group and representations, we
find that the induced effective potential $V(\alpha)$ has
minima at $\alpha\sim 0.2$, leading to too-low values for $1/R$.
The Higgs mass $m_H=(g_4R/2)\sqrt{V^{\prime\prime}(\alpha)}$, which is
further suppressed by a gauge loop factor, tends consequently to be
too low as well. Finally, the top mass $m_{\rm top}$ cannot be
adjusted to a high-enough value.  A possible way of improving the
situation is to add to the model new heavy bulk fermions that do not
couple to the SM matter fields but transform in large representations
of the gauge group, in such a way as to modify
$V(\alpha)$ and obtain minima for much lower $\alpha$. For
representations of rank $r \gg 1$, one can obtain minima at values as
low as $\alpha \sim 1/r$, with a $V^{\prime\prime}(\alpha)$ at the
minimum that rapidly grows with $r$. This helps considerably in
increasing $1/R$ and $m_H$, and to some extent $m_{\rm top}$. However,
fields with high rank $r$ induce electroweak quantum corrections that are
enhanced by large group-theoretical factors, thereby lowering the
cut-off and the predictive power of the theory.

A simple and natural generalization of our minimal set-up consists in
introducing additional localized kinetic terms for gauge fields
\cite{Carena:2002me}\footnote{We thank R.~Rattazzi for suggesting
  this possibility to us.}. These have the effect of increasing the bulk
gauge coupling and to distort in a non-trivial way the wave functions
and mass spectrum of the Kaluza--Klein (KK) modes of the gauge bosons.
They also modify the gauge contribution to the effective potential,
which we explicitly compute for arbitrary values of the localized
couplings. It turns out that for sufficiently large values of the
latter, where the analysis simplifies, it is possible to get values of
$\alpha$ leading to acceptable values of $1/R$; at the same time,
$m_H$ is significantly increased and $m_{\rm top}$ can be reproduced.
We show, however, that the simultaneous presence of localized
gauge kinetic terms and a Wilson-line symmetry breaking leads in
general to an unwanted deformation of the electroweak sector of the
theory. The main point is that the gauge interactions are no longer
universal and the masses of the lowest KK gauge bosons are
deformed.  This implies, among other things, that the tree-level value
of the $\rho$ parameter departs from $1$ and hence imposes generically
very severe bounds on the size of localized gauge kinetic terms. These
could strongly constrain model building even if they are only
radiatively generated, although it is not excluded that large values
can be tolerated in special models with new symmetries. Finally, the
scale at which electroweak interactions become strong is again
lowered, so that the theory tends to become less predictive.

The paper is organized as follows. In section 2 we present a basic
model that captures all the essential features we want to exploit. In
section 3, the non-local Yukawa couplings between boundary fermions
and the Higgs field are computed. In section 4 we compute the one-loop
effective potential for the Higgs field. In section 5 we construct a
prototype 5D model and discuss its main properties. In section 6, the
effects of localized gauge kinetic terms are studied. Finally, section
7 is devoted to some general conclusions, and some technical details
are collected in the appendix.

\section{The basic construction}

The simplest framework allowing an implementation of the Higgs field
as the internal component of a gauge field is a 5D gauge theory with
gauge group $G=SU(3)$ on an $S^1/\Z_2$ orbifold \cite{Antoniadis:1993jp}.
The $\Z_2$ orbifold projection is embedded in the gauge group
through the matrix
\begin{equation}
P = e^{i\pi \lambda_3} =
\left(
\matrix{
-1 \a 0 \a 0 \cr
 0 \a -1\;\;\, \a 0 \cr
 0 \a  0 \a 1 \cr}\;
\right)\;,\label{Rtwist}
\end{equation}
where $\lambda_a$ are the standard $SU(3)$ Gell-Mann matrices,
normalized as ${\rm Tr}\, \lambda_a \lambda_b = 2 \delta_{ab}$,
so that $A_M = A_M^a \lambda_a/2$. The group $G$ is broken in 4D
to the commutant $H=SU(2) \times U(1)$ of the projection $P$.
The massless 4D fields are the gauge bosons $A_\mu^a$ in the
adjoint of $H$ and a charged scalar doublet, arising from the
internal components $A_5^a$ of the gauge field. A VEV for $A_5$
induces an additional spontaneous symmetry breaking to
$E=\hat U(1)$, generated by $\hat A_\mu= (A_\mu^8 + \sqrt{3} A_\mu^3)/2$.
Using the residual $H$ symmetry, it is always possible to align
$\langle A_5 \rangle$  along the $\lambda_7$ component, corresponding
to the down component of the doublet, and take
\begin{equation}
\langle A_5^a \rangle = \frac {2\alpha}{g_5 R}\,\delta^{a7} \;.
\label{vev}
\end{equation}
Identifying $H$ with the electroweak gauge group, $E$ with the
electromagnetic group, and the zero-mode of $A_{5}$ with the
Higgs field $H$, the above construction gives a description of
electroweak symmetry breaking.
In this minimal version, the weak mixing angle $\theta_W$ turns out
to be too large, $\theta_W=\pi/3$, but this problem can be solved
by starting with a different unified group $G$, as we will see.
Taking into account the normalization of the zero-mode (see the appendix),
(\ref{vev}) corresponds to a VEV for the neutral component
of the Higgs doublet $H$ equal to  $2\alpha/(g_4 R)$, with
\be
g_4 = \frac {g_5}{\sqrt{2 \pi R}} \;.
\ee

It is well known from \cite{hos,Haba:2002py} that a VEV for $A_5$
induces a Wilson line which is equivalent to a Scherk--Schwarz twist,
the two situations being related through a non-periodic gauge
transformation. The twist matrix $T(\alpha)$ satisfies the consistency
condition $T P T = P$ \cite{consistency}; for the choice (\ref{vev}),
it is given by
\begin{equation}
T(\alpha) = e^{2i\pi \alpha \lambda_7} =
\left(\;\matrix{
1 \a 0 \a 0 \cr
0 \a \cos 2\pi\alpha \a \sin 2\pi\alpha \cr
0 \a -\sin 2\pi\alpha \a \cos 2\pi\alpha \cr}\;
\right)\;.
\label{Ttwist}
\end{equation}
The orbifold projection represents an explicit symmetry breaking of
$G$ to $H$, the masses of the fields in $G/H$ being of order $1/R$,
whereas the Scherk-Schwarz twist corresponds to a spontaneous symmetry
breaking of $H$ to $E$, the masses of the fields in $H/E$ being of
order $\alpha/R$. This situation is equivalent to an $S^1/(\Z_2\times
\Z_2^\prime)$ orbifold with two non-commuting projections $P$ and
$P^\prime(\alpha) = T(\alpha) P = T(\alpha/2) P T(-\alpha/2)$, and
radius $2R$. From this point of view, there are two perfectly
symmetric projections, the first breaking $G$ to $H$ with fixed-points
$y = 0,2\pi R$, and the second $G$ to $H^\prime$ with fixed-points $y
= \pi R,3 \pi R$. Together, they break $G$ to $E = H \cap H^\prime$ in
a non-local way: the two subgroups $H$ and $H^\prime$ are isomorphic,
but their embeddings in $G$ form an angle $\alpha$.  The most general
field content consists of bulk fields in representations of $G$, and
boundary fields in representations of the subgroup $H$ or $H^\prime$
surviving at the fixed-point where they are localized (see e.g.
\cite{HH}).

In the following, we will adopt the perspective of the $S^1/\Z_2$
orbifold with a Wilson line breaking, and work on the fundamental
interval $y \in [0,\pi R]$. The brane fields of our basic construction
consist of a left-handed fermion doublet $Q_L=(u_L,d_L)^T$ and two
right-handed fermion singlets $u_R$ and $d_R$ of $H=SU(2)\times U(1)$;
these fields can also be described by their charge conjugates
$Q_R^c=(d^c_R,-u^c_R)^T$, $d^c_L$ and $-u^c_L$. We will assume that
the doublet and the two singlet fields are located respectively at
positions $y_1$ and $y_2$, equal to either $0$ or $\pi R$. The bulk
fields are, in addition to the gauge fields $A_M$, one pair of
fermions $\Psi_a$ and $\tilde\Psi_a$ with opposite $\Z_2$ parities for
each type of matter field $a=u,d$; we take these in the symmetric and
the fundamental representations of $G=SU(3)$ for $a=u$ and $a=d$
respectively.

The parity assignments for the bulk fermions allow for bulk mass terms
$M_a$ mixing $\Psi_a$ and $\tilde \Psi_a$, as well as boundary
couplings $e_{1,2}^a$ with mass dimension $1/2$ mixing the
bulk fermion $\Psi_a$ to the boundary fermion $a$. Denoting the
doublet and singlet components arising from the decomposition of the
bulk fermion $\Psi_{a}$ under $G \rightarrow H$ (see the appendix)
by $\psi_a$ and $\chi_a$ respectively, the complete Lagrangian for
matter fields is then given by the following
expression\footnote{$\Dslashn_4$ and $\Dslashn_5$ denote the $4D$ and
  $5D$ covariant derivatives. Defining the Hermitian matrix $\gamma_5
  = i\,\gamma_4$, they are related by $\Dslashn_5 = \Dslashn_4 +
  i\gamma_5 D_5$.}:
\begin{eqnarray}
\label{Lagferm}
\mathcal{L}_{\rm mat}\a=\a \raisebox{0pt}{$\sum_{a}$}
\Big[\bar \Psi_a \,i \Dslash_5\, \Psi_a
+ \tilde{\bar \Psi}_a \,i \Dslash_5\, \tilde \Psi_a
+ \Big(\bar \Psi_a M_a \tilde \Psi_a +\mathrm{h.c.} \Big)\Big] \nn \\
\a\;\a +\,\delta(y-y_1) \Big[\bar Q_L \,i \Dslash_4\, Q_L
+ \Big(e_1^d \bar Q_L \psi_{d} + e_1^u \bar Q_R^c \psi_{u}
+ \mathrm{h.c.}\Big) \Big] \nn \\
\a\;\a +\,\delta(y-y_2) \Big[\bar d_R \,i \Dslash_4\, d_R
+ \bar u_R \,i \Dslash_4\, u_R
+ \Big(e_2^d \bar d_R \chi_{d} + e_2^u \bar u_L^c \chi_{u}
+ \mathrm{h.c.}\Big)\Big] \,. \nn
\end{eqnarray}
All bulk fermion modes are massive and, neglecting the
bulk-to-boundary couplings, their mass spectrum is given by $M_{a,n} =
\pm \sqrt{m_n^{2} + M_a^2}$, where $m_n=n/R$.  After the spontaneous
symmetry breaking induced by (\ref{vev}), a new basis has to be defined
for the bulk fermion modes in which they have diagonal mass
terms, with a shift in the KK masses $m_{n} \rightarrow
m_{n}(\alpha)$.  The procedure is outlined in the appendix. The new
fields $\Psi^{(i)}$ and $\tilde \Psi^{(i)}$ defining this basis are
given by eqs.~(\ref{fundredef}) and (\ref{symmredef}), and have KK
masses $m_{n}^{(i)}=(n+q^{(i)}\alpha)/R$, with $q^{(i)}$ being an
integer charge, defined by eqs.~(\ref{massefun}) and (\ref{massesym});
$i=1,2$ for the fundamental representation and $i=1,2,3,4$ for the
symmetric one. Similarly, a new basis has to be defined for the
gauge-field modes to diagonalize their mass terms. The new fields
$A_M^{(i)}$ and their KK masses are defined as in eqs.~(\ref{adjredef})
and (\ref{masseadj}), where the field $\Psi^\pm$ are respectively identified
with the gauge field components $A_\mu$ and $A_5$. More precisely,
the complex gauge field $A_\mu^{(1)}$ with charge $q^{(1)} = 1$
($\Psi_0^{+(1)}$) is identified with the $W$ boson, the real field
$A_\mu^{(2)}$ with $q^{(2)} = 2$ ($\Psi_0^{+(2)}$) with the $Z$ boson and
the neutral field $A_\mu^{(0)}$ ($\Psi^{+(3)}_0$) with the photon.
Similarly, the scalar field $A_5^{(0)}$ ($\Psi_0^{-(4)}$) is identified
with the component of the Higgs field $H$ that is left over after the
spontaneous symmetry breaking. Using the fact that $\sec \theta_W = 2$,
the masses of the $W$ and $Z$ fields can be written
as
\begin{equation}
m_W = \frac{\alpha}R \;,\;\;m_Z = \frac{\alpha}R\, \sec \theta_W \;.
\label{MW}
\end{equation}
The Higgs mass is radiatively induced after the electroweak symmetry
breaking and depends on the second derivative of the potential
evaluated at the minimum:
\begin{equation}
m_H = \frac{g_4 R}2 \, \sqrt{V^{\prime\prime}(\alpha)} \;.
\label{MH}
\end{equation}

In the following, it will be convenient to take the size $\pi R$ of
the orbifold as reference length scale and use it to define
dimensionless quantities. In particular, it will be useful to
introduce the parameters $\lambda^a = \pi R M_a$ and $\epsilon_i^a =
\sqrt{\pi R/2} e_i^a$, and the integer $\delta = (\pi R)^{-1}|y_1-y_2|$
parametrizing the distance between the location of left- and right-handed
fields.

\section{Induced couplings for the boundary fermions}

The heavy bulk fields couple to boundary fermions, and can therefore
induce gauge-invariant non-local couplings mixing matter fermions
localized at the fixed-points and the field $A_5$ through Wilson lines
\cite{Csaki:2002ur}. Since the VEV (\ref{vev}) for $A_5$ mixes the
modes of different components of the bulk fermions, it is convenient
to use the fields $\Psi^{(i)}$ and $\tilde \Psi^{(i)}$ defined before
and to group them into a two-vector $(\Psi^{(i)},\tilde
\Psi^{(i)})^T$. In this notation, the kinetic term of the $n$-th KK
mode of each component of the bulk fermions is encoded in the
following two-by-two matrix in momentum space:
\begin{equation}
K^{\Psi} = \left(\,\matrix{
\pslash - m_n \a M \cr
M \a \pslash + m_n \cr}
\right)\;.
\label{KPsi}
\end{equation}
The tree-level propagator is obtained by inverting this matrix, and is
given by
\begin{equation}
\Delta^{\Psi} = \frac i{p^2 - m_n^{2}- M^2}
\left(\,\matrix{
\pslash + m_n \a - M \cr
-M \a \pslash - m_n \cr}\,
\right) \;.
\label{prop}
\end{equation}

The couplings between bulk and boundary fields can be easily rewritten
in terms of the new fields as well; the new doublet and singlet
components under the decomposition $G \rightarrow H$ can be easily
read off from eqs.~(\ref{wavefun}) and (\ref{wavesym}). The only bulk
field that couples both to the left-handed and right-handed boundary
fields, and can therefore induce mass terms, is always $\Psi^{(2)}$.
The other components of the bulk fermions couple only separately to
left-handed and right-handed fields; they can therefore induce only
wave-function corrections. For each bulk-to-boundary coupling, there is
a wave-function factor $\xi_{i,n}$ for the $n$-th mode, which is equal
to $1$ if $y_i=0$ and $(-1)^n$ if $y_i=\pi R$. The interaction
Lagrangian reads
\begin{eqnarray}
{\cal L}_{\rm int} \a=\a
\frac 1{\pi R} \sum_{n=-\infty}^\infty
\Big[\epsilon_1^u\, \xi_{1,n} \Big(\bar d_R^c \Psi_{u,n}^{(1)}
- \bar u_R^c \Psi_{u,n}^{(2)}\Big)
- \frac {\epsilon_2^u}{\sqrt{2}}\, \xi_{2,n} \bar u_L^c
\Big(\Psi_{u,n}^{(2)} + \eta_n \Psi_{u,n}^{(4)}\Big) \nn \\
\a\;\a \hspace{58pt} +\,\epsilon_1^d\, \xi_{1,n}
\Big(\eta_n \bar u_L \Psi_{d,n}^{(1)} + \bar d_L \Psi_{d,n}^{(2)}\Big)
+ \epsilon_2^d\, \xi_{2,n} \bar d_R \Psi_{d,n}^{(2)} +
{\rm h.c.} \Big] \;,\qquad
\label{Lint}
\end{eqnarray}
where the factor $\eta_n$ is defined to be 1 for $n=0$ and
$1/\sqrt{2}$ for $n\neq 0$.  The bulk fermions can be disentangled
from the boundary fermions by completing the squares in the kinetic
operator and performing an appropriate redefinition of the bulk
fields. This generates corrections to the kinetic operators $K^u$ and
$K^d$ for the boundary fields $u=u_L+u_R$ and $d=d_L+d_R$, originally
given just by $\pslash$, which correspond diagrammatically to the
exchange of bulk fermions. These contributions have the structure
$\sum_n (e\, \xi_{n} P_{L,R}) \Delta_{n}^\Psi (e\, \xi_{n} P_{R,L})$,
where $P_{L,R}$ are chiral projectors and $\Delta_{n}^\Psi$ is given
by the first entry of (\ref{prop}) with $m_{n}$ as in
eqs.~(\ref{massefun}) and (\ref{massesym}). To compute the infinite
sums, it is convenient to go to Euclidean space, and define the
dimensionless momentum variables $x = \pi R p$ and $x^a = \pi R
\sqrt{p^2 + M_a^2}$, as well as the basic functions $f_\delta$ given
by
\begin{eqnarray}
f_0(x,\alpha) \a=\a \sum_{n=-\infty}^\infty \frac 1{x + i \pi (n+\alpha)}
= \coth (x + i \pi \alpha) \;, \label{f0} \\
f_1(x,\alpha) \a=\a \sum_{n=-\infty}^\infty \frac {(-1)^n}{x +
  i \pi (n+\alpha)}
= \sinh^{-1} (x + i \pi \alpha) \;. \label{f1}
\end{eqnarray}
The functions $f_\delta(x,\alpha)$ are related to the propagation of
bulk fields between two fixed-points separated by a distance $\delta\,\pi R$.
This is particularly clear from their Taylor expansion, which takes the
simple form
\begin{equation}
f_\delta(x,\alpha) =
\sum_{k=-\infty}^\infty e^{-|2k+\delta|(x + i \pi \alpha)}
\label{f}\;.
\end{equation}

After a straightforward computation, the total effective actions for
the boundary fields are found to be given by the following
expressions:
\begin{eqnarray}
K^u \a=\a \pslash\,{\rm Re}\Bigg[1
+ \frac {\epsilon_1^{d\s 2}\!}{x^d} P_L\s f_0(x^d\!,0)
+ \frac {\epsilon_2^{u\s 2}\!}{2x^u} P_R\s f_0(x^u\!,0)
+ \Big(\frac {\epsilon_1^{u\s 2}\!\!}{x^u} P_L
+ \frac {\epsilon_2^{u\s 2}\!}{2x^u} P_R \Big) f_0(x^u\!,2\alpha)\Bigg] \nn \\
\a\;\a +\,\frac {i}{\pi R}\,{\rm Im} \,
\Bigg[\frac {\epsilon_1^u \epsilon_2^u}{\sqrt{2}}
f_\delta(x^u\!,2\alpha)\Bigg] \;,
\label{Ku} \\
K^d \a=\a \pslash\,{\rm Re}\Bigg[1
+ \Big(\frac{\epsilon_1^{d\s 2}}{x^2} P_L
+ \frac {\epsilon_2^{d\s 2}\!}{x^d} P_R \Big) f_0(x^d\!,\alpha)
+ \frac{\epsilon_1^{u\s 2}\!}{x^u} P_L\s f_0(x^u\!,\alpha) \Bigg] \nn \\
\a\;\a +\,\frac {i}{\pi R}\,{\rm Im} \,\Bigg[\epsilon_1^d \epsilon_2^d\,
f_\delta(x^d\!,\alpha)\Bigg] \;.
\label{Kd}
\end{eqnarray}
{}From the above expressions, we see that an infinite set of
non-renormalizable interactions involving the Higgs field are
generated together with the standard Yukawas. The new interactions
have small effects in physical processes as long as the radius $R$ is
small enough, but they all contribute comparably to spontaneous
symmetry breaking, whose effect is not only to induce masses $m^u$ and
$m^d$ mixing $d_L$ with $d_R$ and $u_L$ with $u_R$, but also to
generate different wave-function corrections $Z_{1,2}^u$ and
$Z_{1,2}^d$ for $u_{L,R}$ and $d_{L,R}$. Moreover, all these
quantities are momentum-dependent.

To be precise, the boundary fields have an effect on the spectrum of
the bulk fields as well, which in general cannot be neglected, and the
mass eigenstates are mixtures of bulk and boundary states. To find
the exact spectrum of fermions, one would have to diagonalize the full
kinetic operator for the entangled bulk and boundary fermions.
Assuming however that the physical mass induced for the boundary
fields is much smaller than the masses of the bulk fields, one can
neglect the distortion on the spectrum of the latter and use the free
kinetic operator (\ref{KPsi}) for bulk fermions and the results
(\ref{Ku}) and (\ref{Kd}) for the boundary fields. In this
approximation, the spectrum of bulk fields is unchanged and the mass
of the boundary fields is obtained by looking at the zeros (\ref{Ku})
and (\ref{Kd}).  Notice that the momentum dependence in the latter
can be safely neglected within the adopted approximation. The
masses $m^a$ and the wave-functions $Z_{1,2}^a$ for left and right
components then reduce to $\alpha$-dependent parameters given by
\begin{eqnarray}
m^u \a=\a \frac {\epsilon_1^u \epsilon_2^u}{\sqrt{2} \pi R}\,
{\rm Im}\s f_\delta(\lambda^u\!,2\alpha)
\label{mu} \;, \\
m^d \a=\a \frac {\epsilon_1^d \epsilon_2^d}{\pi R}\,
{\rm Im}\s f_\delta(\lambda^d\!,\alpha)
\label{md} \;, \\
Z_i^u \a=\a 1
+ \delta_{i1} \frac {\epsilon_1^{d\s 2}\!}{\lambda^d}\,
{\rm Re}\s f_0(\lambda^d\!,0)
+ \delta_{i2}\, \frac {\epsilon_2^{u\s 2}\!}{2\lambda^u}\,
{\rm Re}\s f_0(\lambda^u\!,0)
+ \frac {\epsilon_i^{u\s 2}\!}{2^{\delta_{i2}}\lambda^u}\,
{\rm Re}\s f_0(\lambda^u\!,2\alpha)
\label{Ziu} \;, \qquad \\
Z_i^d \a=\a 1
+ \frac {\epsilon_i^{d\s 2}\!}{\lambda^d}\,
{\rm Re}\s f_0(\lambda^d\!,\alpha)
+ \delta_{i1} \frac {\epsilon_1^{u\s 2}\!}{\lambda^u}\,
{\rm Re}\s f_0(\lambda^u\!,\alpha)
\label{Zid} \;.
\end{eqnarray}
Finally, the physical masses after symmetry breaking are obtained by
rescaling the fields to canonically normalize them; they are given by
\begin{equation}
m_{\rm phys}^a = \frac{m^a}{\sqrt{Z_1^a Z_2^a}}\,.
\label{mfis}
\end{equation}

The arguments of \cite{ms3} suggest that the effective actions induced
for the boundary fields should be non-local from the $5D$ point of
view and generated when $\langle A_5 \rangle$ acquires the VEV
(\ref{vev}) from operators involving Wilson lines
\begin{equation}
W_{n}(A_5) = {\cal P} \exp \Big\{i g_5 \int_0^{n \pi R} \!\!\!
dy A_{5} \Big\}\,.
\label{WL}
\end{equation}
Indeed, the $k$-th term in the series expansion (\ref{f}) corresponds
to the components of $W_{2|n|}(\langle A_5 \rangle) = T(|n|\alpha)$
with a winding number $n = 2k+\delta$; this connects the two
boundaries through a path that winds $k$ times around the internal
circle, with total length $|n|\pi R$.  Since the boundary fields break
explicitly $G$ to $H$, the effective action will be only $H$-invariant
and involve the various components of the decomposition of (\ref{WL}).
However, it is nevertheless convenient to embed the boundary fields
$Q$ and $q$ into $G$ representations of the same type as those of the
bulk fields, completing them with vanishing components. This
description of the boundary fields is similar to the one that would
emerge for the boundary values of bulk fields with $+$ and $-$
parities. More precisely, the boundary fields $Q$ and $q$ can be
embedded into fundamental representations $Q_F$ and $q_F$ or symmetric
representations $Q_S$ and $q_S$ of $SU(3)$ as follows:
\begin{eqnarray}
Q_F =
\left(\matrix{
u_L \cr d_L \cr 0 \cr
}\right) \,,\;\,
d_F =
\left(\matrix{
0 \cr 0 \cr d_R \cr
}\right) \,,\;\,
Q_S^c = \frac 1{\sqrt{2}}
\left(\matrix{
0 \a 0 \a d_R^c \cr
0 \a 0 \a \!\mbox{-}u_R^c\! \cr
d_R^c \a \mbox{-}u_R^c \a 0 \cr
}\right) \,,\;\,
u_S^c =
\left(\matrix{
0\; \a 0 \a 0 \cr 0\; \a 0 \a 0 \cr
0\; \a 0 \a \mbox{-}u_L^c \cr
}\right) \,.
\label{emb}
\end{eqnarray}
In this notation, the couplings between boundary and bulk
fermions can be obtained simply by taking traces of products of
eqs.~(\ref{emb}) with eqs.~(\ref{wavefun}) and (\ref{wavesym}).

Using the embeddings (\ref{emb}), one can easily verify that the
leading part of the effective action in a derivative expansion is
given by the sum of the original action ${\cal L}_0 = \bar Q i \dslash
Q + \bar u i \dslash u + \bar d i \dslash d$ and the following
non-local interactions ${\cal L}_F$ and ${\cal L}_S$ induced by the
bulk fermions in the fundamental and symmetric representations
respectively (going back to Minkowski space):
\begin{eqnarray}
{\cal L}_{F} \a=\a
\sum_k e^{-|2k|\lambda^d} \Bigg[
\frac {\epsilon_1^{d\s 2}\!}{\lambda^d} \bar Q_F W_{|2k|} i \dslash Q_F
+ \frac {\epsilon_2^{d\s 2}\!}{\lambda^d} \bar d_F W_{|2k|} i \dslash d_F
\Bigg]\nn \\
\a\;\a + \sum_k e^{-|2k+\delta|\lambda^d}
\frac {\epsilon_1^d \epsilon_2^d}{\pi R}
\Bigg[\bar Q_F W_{|2k+\delta|} d_F + {\rm h.c.} \Bigg] \;,
\label{meffnlfun} \\
{\cal L}_{S} \a=\a
\sum_k e^{-|2k|\lambda^u} \Bigg[
\frac {\epsilon_1^{u\s 2}\!}{\lambda^u} {\rm Tr}\,W^T_{|2k|}
\bar Q_S^c W_{|2k|} i \dslash Q_S^c
+ \frac {\epsilon_2^{u\s 2}\!}{\lambda^u} {\rm Tr}\,W_{|2k|}^T
\bar u_S^c W_{|2k|} i \dslash u_S^c \Bigg] \nn \\
\a\;\a +\,\sum_k e^{-|2k+\delta|\lambda^u}
\frac {\epsilon_1^u \epsilon_2^u}{\pi R}
{\rm Tr}\,\Bigg[W_{|2k+\delta|}^T \bar Q_S^c W_{|2k+\delta|} u_S^c
+ {\rm h.c.} \Bigg] \;.
\label{meffnlsym}
\end{eqnarray}

We conclude that it is possible to induce generic Yukawa couplings for
boundary fields through non-local operators involving Wilson lines
that connect the two boundaries and wind around the orbifold an
arbitrary number of times. The resulting physical mass $m_{\rm phys}^a$
of the boundary fields is exponentially sensitive to the parameter
$\lambda^a$ governing the bulk masses. Notice also that, because of the
wave-function rescaling, the value of $m_{\rm phys}^a$ given by
(\ref{mfis}) cannot be made arbitrarily large by increasing the
values of the boundary couplings $\epsilon_i^a$. Indeed, for
$\epsilon_i^a \gg 1$, $m_{\rm phys}^a$ quickly saturates to a value
depending only on ratios of these parameters but not on their overall
size.

\section{One-loop effective potential for the Higgs}
\label{sec:radiative}

The field $A_5$ couples only to the gauge fields and to the two bulk
fermions. Its radiatively induced potential thus depends only
indirectly on the boundary couplings through diagrams in which the
virtual bulk fermions temporarily switch to a virtual boundary
fermion. The total potential is therefore the sum of a universal gauge
contribution and a parameter-dependent contribution coming from the
fermions.

The contribution of the fermions is obtained by summing up all
possible one-loop diagrams of bulk fermions dressed by an arbitrary
number of external $A_5$ lines and insertions of boundary couplings.
Since bulk interactions conserve the KK momentum, whereas boundary
interactions do not, it is convenient to separately resum diagrams
with no insertion of boundary interactions and diagrams with an
arbitrary but non-zero number of these. The first piece corresponds to
the contribution of bulk fields in the absence of boundary couplings,
with kinetic operator (\ref{KPsi}), whereas the second can be reinterpreted
as the contribution of boundary fermions propagating with the
effective kinetic operators (\ref{Ku}) and (\ref{Kd}) induced by the
insertions of mixings with the bulk field. This decomposition
corresponds precisely to the shift performed in the last section to
disentangle bulk and boundary fermions. In this case, however, this
leads to an exact result, because both bulk and boundary fields are
integrated out.

The bulk fields $\Psi_{a}^{(i)}$ couple to the gauge field $A_5$ in a
diagonal way, through the shift induced in their KK mass by the
minimal coupling. For each mode one has $m^{(i)} = (n + q^{(i)}
\alpha)/R$, where $q^{(i)}$ is the charge of the mode as specified by
eqs.~(\ref{massefun}) and (\ref{massesym}). In total, there are only
three kinds of modes with non-vanishing charges, two with $q=1$ and one
with $q=2$. The contribution to the effective action from a pair of
such modes for the $\Psi^{(i)}$ and $\tilde \Psi^{(i)}$ fields with
given charge $q$ reads $\Gamma^{\Psi}(q\alpha) = - {\rm ln}\,{\rm
det}[K^{\Psi}(q\alpha)]$, where $K^{\Psi}$ is the Euclidean
continuation of (\ref{KPsi}). The determinant of $K^{\Psi}$ as a
two-by-two matrix yields $p^2 + m_n^2 + M^2$.  The determinant in the
space of KK modes then reduces to an infinite product over these
factors, which yields an irrelevant $\alpha$-independent divergence
plus a finite $\alpha$-dependent function. Finally, the determinant
over spinor indices yields a trivial factor of $4$. For a pair of
modes with charge $q$, one then finds
\begin{eqnarray}
V_{\Psi_a}(q\alpha) \a=\a \frac{-1}{2\pi^6 R^4} \int_0^\infty \!\!\! dx\,x^3\,
{\rm ln}\,\Big|f_1(x^a,q\alpha)\Big|^{-2} \nn \\
\a=\a \frac{3}{8 \pi^6 R^4}\sum_{k=1}^\infty
\frac {1 + 2 k \lambda^a + 4/3 k^2 \lambda^{a\,2}}{k^5} e^{- 2 k \lambda^a}
\cos (2 q k \pi \alpha) \,.
\label{Va-xCorr}
\end{eqnarray}

The boundary fields $a=u,d$ couple to $A_5$ only through the non-local
Wilson line effective interactions induced by the bulk fermions, and
their kinetic operators $K^a(\alpha)$ in Euclidean space are given by
(\ref{Ku}) and (\ref{Kd}). Their contributions to the effective action
read $\Gamma^{a}(\alpha) = - {\rm ln}\,{\rm det}[K^a(\alpha)]$.  To
evaluate this expression, we start by using the standard trick of
rewriting it in terms of the scalar quantity $K^a K^{a T}$. The
determinant over spinor indices then yields just a factor of $4$, and
dropping irrelevant constant terms one finds:
\begin{eqnarray}
V_{u}(\alpha) \a=\a \frac{-1}{4\pi^6 R^4}
\int_0^\infty \!\!\! dx\,x^3\,
{\rm ln}\,\Bigg[\prod_{i=1}^2{\rm Re} \Big[1
+ \delta_{i1} \frac{\epsilon_1^{d\s 2}\!}{x^d} f_0(x^d\!,0)
+ \delta_{i2} \frac{\epsilon_2^{u\s 2}\!}{2x^u} f_0(x^u\!,0) \nn \\
\a\;\a \hspace{95pt} +\, \frac{\epsilon_i^{u\s 2}\!}{2^{\delta_{i2}}x^u}
f_0(x^u\!,2\alpha) \Big]
+ \prod_{i=1}^2{\rm Im} \Big[\frac {\epsilon_i^{u\s 2}\!}{2^{\delta_{i2}}x}
f_\delta(x^u\!,\alpha)\Big]\Bigg] \,, \label{boundpotu} \hspace{15pt}\\
V_{d}(\alpha) \a=\a \frac{-1}{4\pi^6 R^4}\int_0^\infty \!\!\! dx\,x^3\,
{\rm ln}\,\Bigg[\prod_{i=1}^2{\rm Re} \Big[1
+ \frac{\epsilon_i^{d\s 2}\!}{x^d} f_0(x^d\!,\alpha)
+ \delta_{i1} \frac{\epsilon_1^{u\s 2}\!}{x^u} f_0(x^u\!,\alpha)
\Big] \nn \\ \a\;\a \hspace{95pt}
+ \prod_{i=1}^2{\rm Im} \Big[\frac {\epsilon_i^{d\s 2}\!}{x}
f_\delta(x^d\!,\alpha)\Big]\Bigg] \;. \label{boundpotd}
\end{eqnarray}

The full contribution of bulk and boundary fermions to the one-loop
effective potential is finally given by
\begin{equation}
V_{f}(\alpha) = V_{\Psi_u} (\alpha) + V_{\Psi_u} (2\alpha)
+  V_{\Psi_d} (\alpha) + V_{u}(\alpha) + V_{d}(\alpha) \;.
\label{Vftot}
\end{equation}
As expected, the result is finite, thanks to the exponentially soft UV
behaviour of the functions $f_\delta$. Notice also that the argument of
the logarithm in the boundary contributions can be rewritten as the
determinant of a two-by-two matrix given by the identity $\delta_{ij}$,
plus a matrix $\Delta_{ij}$ encoding interactions between the
fixed-points located at $y_i$ and $y_j$ and involving the function
$f_{\delta_{ij}}$ with $\delta_{ij} = (\pi R)^{-1} |y_i - y_j|$.

It is worth noting that the result (\ref{Vftot}) contains indirect
information on the exact spectrum of the bulk and boundary fermions,
which allows in fact to probe the accuracy of (\ref{mfis}). The
information about the spectrum of eigenvalues $\xi_n = \pi R m_n$ can
be extracted by comparing our result (\ref{Vftot}) with the definition
of the effective action as a trace over all the mass eigenstates of
the logarithm of their free kinetic operators. Turning the sum into a
product inside the logarithm, one would then obtain in this approach a
logarithm argument proportional to $\prod_n (x^2 + \xi_n^2)$, which
has zeros at $x = i \xi_n$. This means that the exact spectrum of
eigenvalues can be obtained by setting the total logarithm argument in
(\ref{Vftot}) to zero. One can verify that the boundary contribution
has poles exactly where the bulk contribution has double zeros
corresponding to the original tower of degenerate bulk modes. Half of
the original zeros remain and correspond to the combination of bulk
fields whose modes are not perturbed. The other orthogonal
combination has a deformed mass spectrum, which is determined, together
with the masses of the boundary fields, by solving the transcendental
equation arising from the argument of the boundary contribution alone.
The latter can be solved numerically, or analytically for the lightest
modes $\xi_0^a$, under the assumption that $\xi_0^a \ll 1$. In that
limit, the momentum dependence can be completely neglected and the
logarithm argument reduces to $Z_1(\alpha) Z_2(\alpha) x^2 + (\pi R
m^a)^2$, which leads to (\ref{mfis}).  This expression is thus valid
as long as $\xi_0^a \ll 1$.  If the latter condition is not satisfied,
the mixing between bulk and boundary modes can have a significant
effect on the mass of the lowest-lying state, which has to be computed
by numerically solving the exact transcendental equation defining the
spectrum.

The contributions to the effective action from gauge and ghost fields
are easily computed \cite{hos,Kubo:2001zc}. Going again to a diagonal
basis, two modes with $q=1$ and one with $q=2$ are found. Each
contributes
\begin{eqnarray}
V_{g}^{A}(q\alpha) \a=\a \frac {3}{16\pi^6 R^4}
\int_0^\infty \!\!\! dx\,x^3\,{\rm ln}\Big|f_1(x,q\alpha)\Big|^{-2} \nn \\
\a=\a-\frac{9}{64 \pi^6 R^4}\sum_{k=1}^\infty \frac{1}{k^5}
\cos (2 q k \pi \alpha) \;,
\label{V5gauge1}
\end{eqnarray}
which gives, in total:
\begin{equation}
V_{g}(\alpha) = 2 V_{g}^A(\alpha) + V_{g}^A(2 \alpha) \;.
\end{equation}

The total one-loop effective potential is given by $V = V_{f} +
V_{g}$.  It satisfies the symmetry property $V(1 \pm \alpha) =
V(\alpha)$.  As expected, the bulk-to-boundary couplings
$\epsilon_i^a$ deform the potential in a non-trivial way; it is
thus interesting to analyse the possible minima one can get in this
case. Notice first that standard 5D bulk fermions in fundamental or
symmetric representations, combined with gauge bosons, can lead to
non-trivial minima for the values $\alpha=0.5$ and $\alpha \sim 0.3$
respectively. In our situation, however, lower values can be obtained,
thanks to the effect of the boundary interactions. This is easily
understood by noticing that for $\alpha \ll 1$ both $V_{a}$ and $V_g$
increase with $\alpha$, whereas $V_{\Psi_a}$ decreases as $\alpha$
increases; the boundary contribution therefore tends to shift the
minimum of $V$ to lower values of $\alpha$. Furthermore, both fermion
contributions are very sensitive to $\lambda^a$ and decrease
exponentially with $\lambda^a$, whereas the dependence on the
brane-to-bulk couplings $\epsilon_i^a$ of the boundary fermion
potential is mild. At fixed $\epsilon_i^a$, the dependence of $V$ on
the $\lambda^a$'s, which we assume to be equal to some common value
$\lambda$ for simplicity, is as follows. For $\lambda=0$, $V_{\Psi_a}$
dominates and we get $\alpha \sim 0.3$, roughly the same value as in
the case of decoupled 5D massless fermions. As $\lambda$ is increased,
$V_{\Psi_a}$ and $V_a$ decrease and the minimum moves to lower values
of $\alpha$, down to $\alpha\sim 0.2$, the precise value depending on
the $\epsilon_i^a$ couplings. When $\lambda$ further increases, $V_g$
eventually dominates and the only minimum that is left is the trivial
one at $\alpha =0$. We have performed a numerical study of $V$ to
determine the lowest values of $\alpha$ that can be achieved in this
setting. We were able to find minima for $\alpha \sim 0.1-0.2$ for a wide
range of the parameters $\lambda^a$ and $\epsilon_i^a$ (see fig. 1).

\begin{figure}[t]
  \begin{center}
    \begin{tabular}{c c}
      \hspace{-10pt}
      \includegraphics[width=0.5\textwidth]{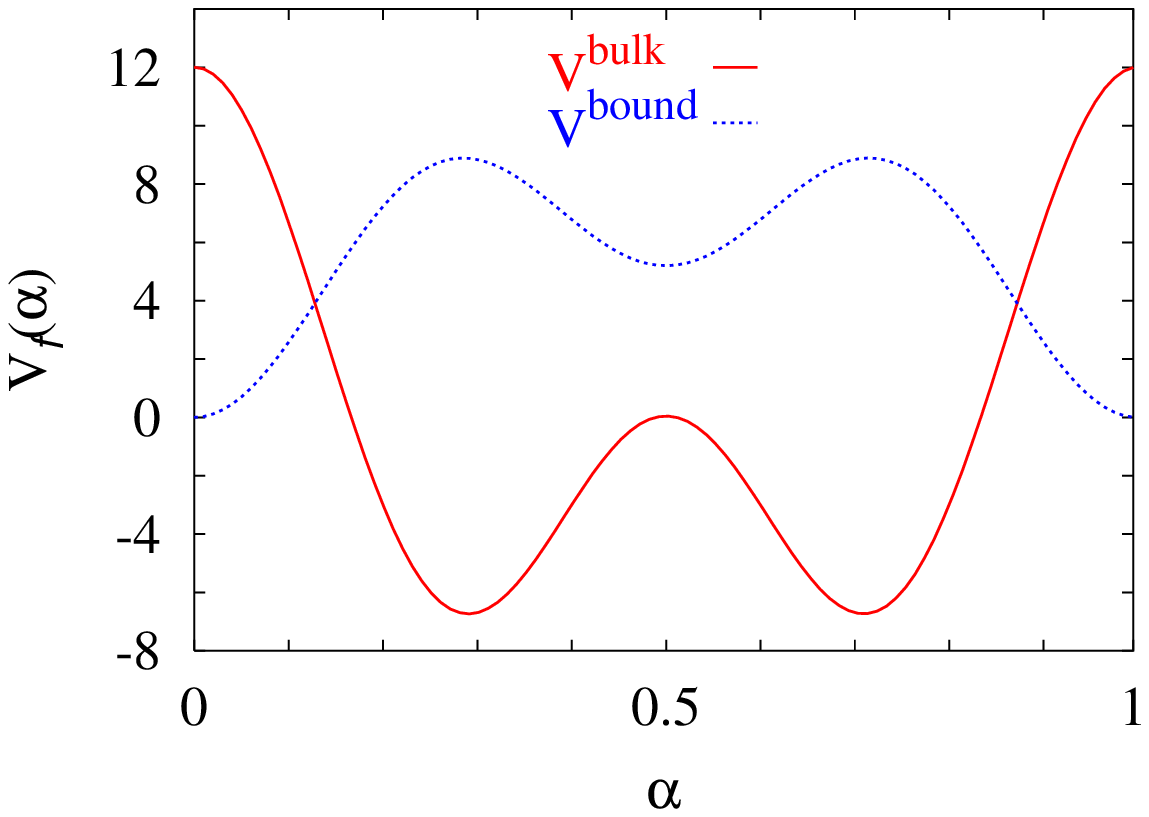} &
      \hspace{-10pt}
      \includegraphics[width=0.5\textwidth]{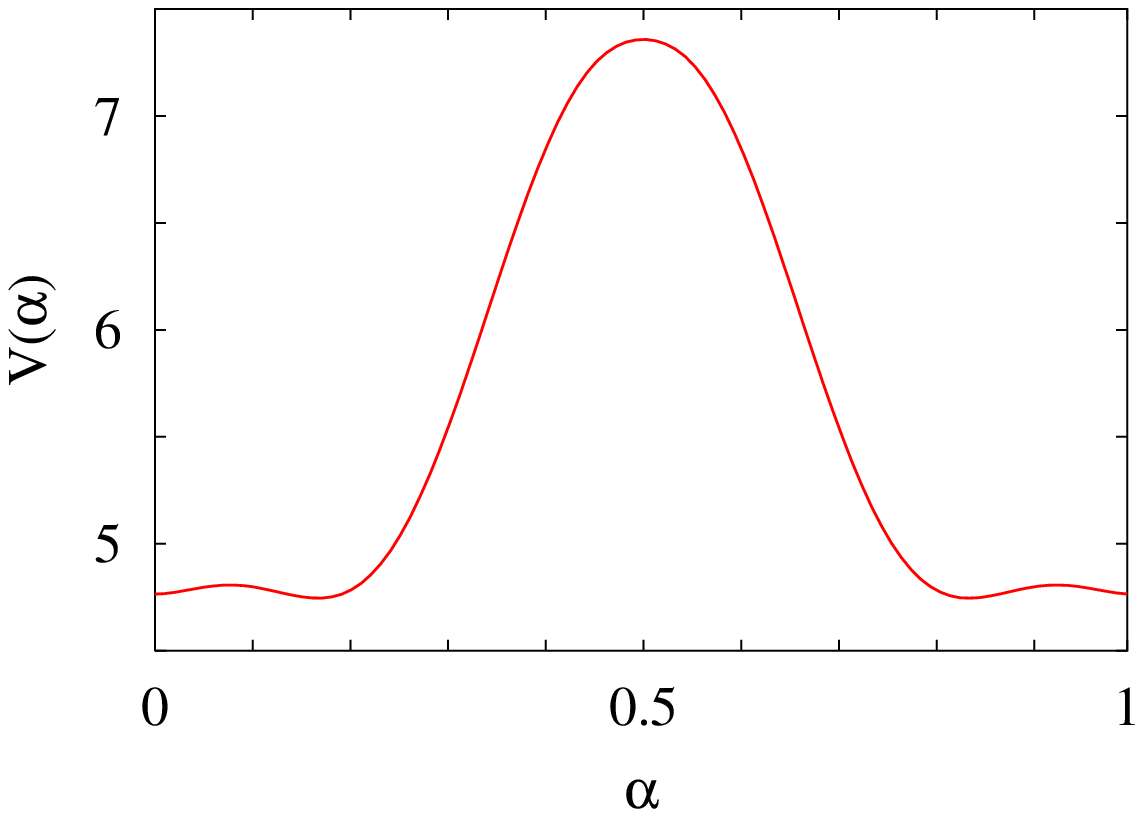} \\
      \hspace{-10pt}
      \includegraphics[width=0.5\textwidth]{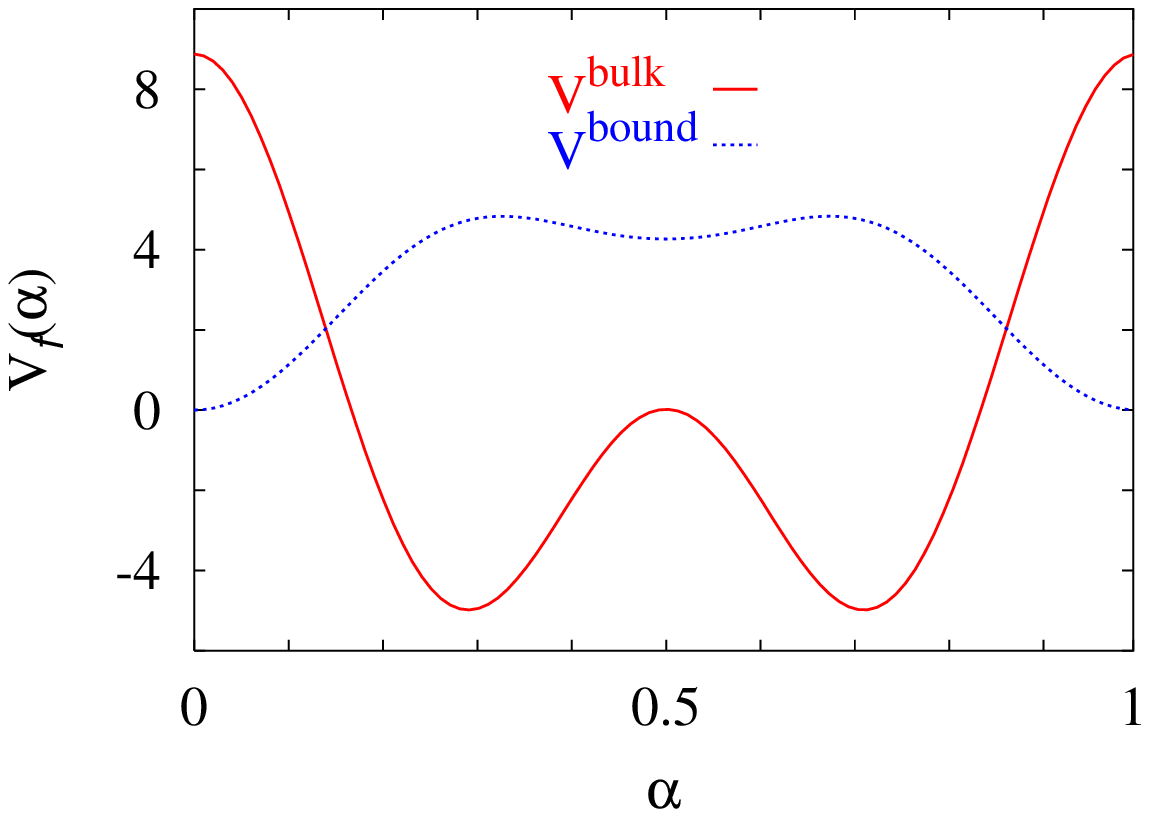} &
      \hspace{-10pt}
      \includegraphics[width=0.5\textwidth]{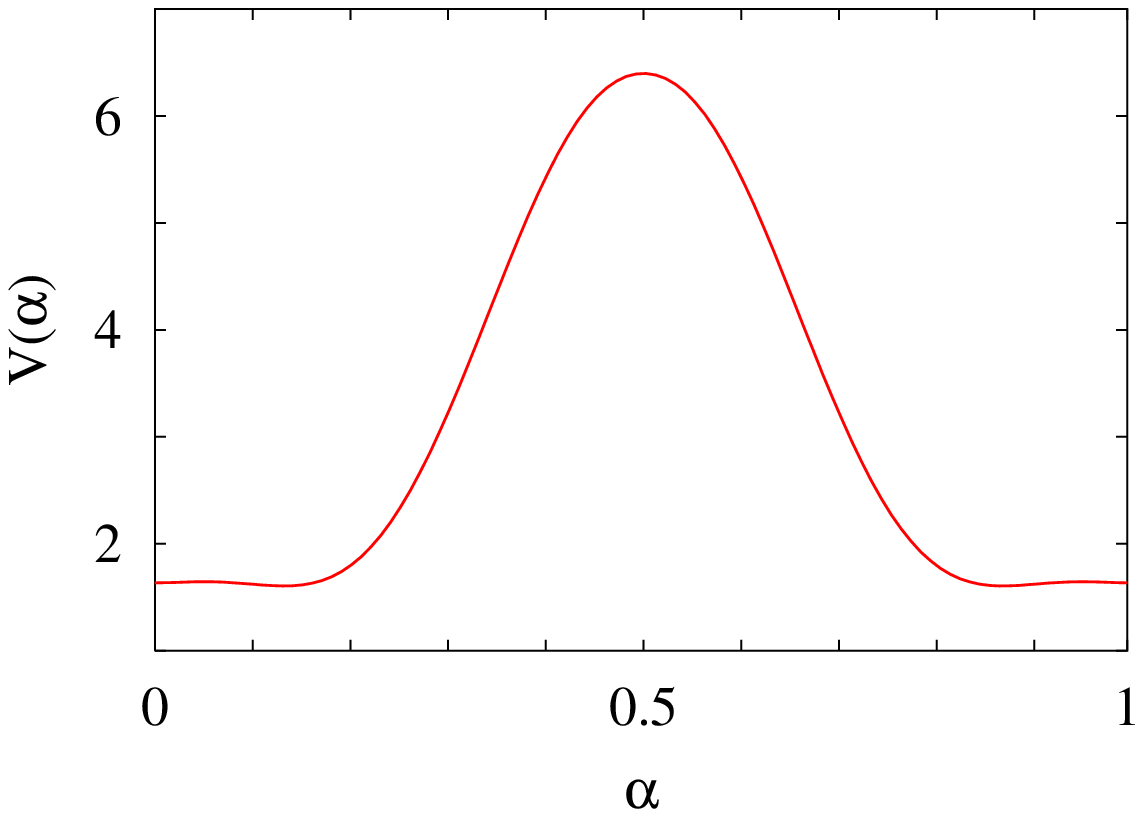} \\
    \end{tabular}
  \end{center}
  \caption{\small
    Different contributions to the effective potential (in
    arbitrary units): the bulk and boundary fermion contributions
    (upper left) and the full potential (upper right) for
    $\lambda=1.57$, $\epsilon_1=3.1$, $\epsilon_2=0.7$ and $\delta =
    0$; the bulk and boundary fermion contributions (lower left) and
    the full potential (lower right) for $\lambda=1.83$,
    $\epsilon_1=6.4$, $\epsilon_2=6.1$ and $\delta = 1$.}
  \label{fig:veff1}
\end{figure}

Let us conclude with a comment on divergences. The above computation
proves that no divergences are induced for the Higgs mass at the one-loop
level, neither on the bulk nor on the boundary. There is actually a
residual gauge invariance at the fixed-points, $A^a_5\to A^a_5 +
\partial_5 \xi^a$, with $a\in G/H$ and $\xi^a$ the corresponding gauge
parameters, which forbids any local boundary mass term for $A^7_{5,0}$
\cite{vonGersdorff}, and this shift symmetry is not broken by the
localized couplings in (\ref{Lagferm}).
This implies that no direct divergence can appear in the potential at any
order in perturbation theory. There will however be linearly divergent
wave-function corrections for the gauge and Higgs fields, which can
substantially influence the physical potential. It should be noticed,
however, that these effects are $G$-symmetric, and $G$-violating
quantities that are only $H$-symmetric are therefore completely
insensitive to them and {\it finite}. Particularly important examples
of such quantities are the ratios $m_W/m_H$ or $m_Z/m_H$, since both the
gauge fields $W,Z$ and the Higgs field $H$ are gauge fields from the
higher-dimensional point of view, and hence receive a common
wave-function correction. At leading order, these ratios depend only
on $\alpha$ and therefore represent predictable quantities for the
model.  They can in principle be influenced by non-renormalizable
operators, so that the accuracy of their leading-order values are
controlled by the effective theory expansion parameter $(\Lambda
R)^{-1}$. However, the very peculiar symmetries constraining $A_5$
do not allow for any higher-dimensional operator that could have
a relevant effect at tree- or one loop-level.

\section{A prototype 5D model}

Having shown how a satisfactory mechanism for electroweak and flavour
symmetry breaking can be achieved, we now turn to the construction of
a simple prototype model in five dimensions. The first concern in
model building is to introduce heavy bulk fermions for each pair of
left- and right-handed SM fermions, so as to obtain all the required
Yukawa couplings for the SM matter (including neutrinos) through the
mechanism explained in section 3.  The strong-interaction sector is
completely factorized and $SU(3)_c$ therefore does not matter. The
true constraint comes from the electroweak-interaction sector and is
fixed by the $SU(2)_L \times U(1)_Y$ quantum numbers of the SM fields;
the Higgs scalar $H$ is a ${\bf 2}_{1/2}$, the quarks $Q_L$, $u_R$, $d_R$
are in the ${\bf 2}_{1/6}$, ${\bf 1}_{2/3}$, ${\bf 1}_{-1/3}$, and the
leptons $L_L$, $l_R$, $\nu_R$ in the ${\bf 2}_{-1/2}$, ${\bf
  1}_{-1}$,${\bf 1}_{0}$.

There are many possibilities for gauge groups unifying electroweak and
Higgs interactions, but we will stick to the basic structure in which
a $SU(3)$ group is broken to $SU(2) \times U(1)$ through a $\Z_2$
orbifold projection.  The decomposition of the simplest $SU(3)$
representations under this projection is as follows: the adjoint
representation decomposes as ${\bf 8} = {\bf 3}_0 \oplus {\bf 2}_{1/2}
\oplus \bar {\bf 2}_{-1/2} \oplus {\bf 1}_{0}$, the fundamental as
${\bf 3} = {\bf 2}_{1/6} \oplus {\bf 1}_{-1/3}$, the symmetric as
${\bf 6} = {\bf 3}_{1/3}\oplus {\bf 2}_{-1/6} \oplus {\bf 1}_{-2/3}$,
and finally the rank-three symmetric as ${\bf 10} = {\bf 4}_{1/2}
\oplus {\bf 3}_{0}\oplus {\bf 2}_{-1/2} \oplus {\bf 1}_{-1}$.

The simplest possibility is to take $G=SU(3)_w$ and $H=SU(2)_L\times
U(1)_Y$.  In this case, bulk fermions in the fundamental, symmetric,
rank-three symmetric and adjoint representations would have the right
charges to couple respectively to the down quarks, conjugate up
quarks, charged leptons and conjugate neutrinos. However, this minimal
choice of the gauge group would lead to too high a weak mixing angle:
$\sin^2\theta_W = 3/4$. A possible cure to this problem consists in
adjusting $\theta_W$ by introducing different localized kinetic
terms for $SU(2)_L$ and $U(1)_Y$ gauge bosons at the fixed-points of
the orbifold. The distortion they cause introduces however other
problems, as we will discuss in the next section.  In addition, the
computation of the Higgs potential is complicated by the presence of
non-universal localized gauge couplings.

An alternative way of tuning $\theta_W$ to a reasonable
value is to add an extra overall $U(1)^\prime$ factor, with coupling
constant $g^\prime$, which remains unaffected by the orbifold
projection, and identify the $U(1)_Y$ hypercharge as the sum of the
$U(1)$ and $U(1)^\prime$ charges after the orbifold projection
$SU(3)_w\times U(1)^\prime \rightarrow SU(2)_L \times U(1) \times
U(1)^\prime$. In this way the gauge field $A_Y$ associated to the
hypercharge and its orthonormal combination $A_X$ are
\begin{equation}
A_Y=\frac{g^\prime A_8 + \sqrt{3} g A^\prime}{\sqrt{3g^2+g^{\prime 2}}}\,, \ \
A_X=\frac{\sqrt{3}g A_8 - g^\prime A^\prime}{\sqrt{3g^2+g^{\prime 2}}}\,,
\label{AX-AY}
\end{equation}
implying that $g_Y = \sqrt{3}g g^\prime/\sqrt{3g^2 + g^{\prime 2}}$ and
thus
\begin{equation}
\sin^2 \theta_W =  \frac{g_Y^2}{g^2 + g_Y^2} =
\frac 3{4 + 3\,g^2/g^{\prime 2}} \;.
\end{equation}
By appropriately choosing the ratio $g/g^\prime$ we can then restore
the correct value of $\sin^2 \theta_W$. After electroweak symmetry
breaking, the following mass term for the gauge fields is induced from
their gauge kinetic term:
\begin{equation}
{\cal L}_{m} = \Big(\frac{\alpha}R\Big)^2
\Bigg[|W|^2 + \frac{1}{2\cos^2 \theta_W}
\Big(Z - \sqrt{3 - 4 \sin^2\theta_W} A_X\Big)^2 \Bigg] \,.
\label{Lm}
\end{equation}
In order for the model to be realistic, the gauge boson $A_X$ must of course
acquire a large mass by some mechanism. As we
shall discuss below, its associated $U(1)_X$ symmetry is actually anomalous,
and is therefore naturally expected to be spontaneously broken at the
cut-off scale, with a mass term
\begin{equation}
{\cal L}_{m}^\prime \simeq \frac 12 \Lambda^2 \big(A_X\big)^2\,.
\label{Lmprime}
\end{equation}
An important consequence of (\ref{Lmprime}) is that the mixing mass
term arising in (\ref{Lm}) between $Z$ and $A_X$ has a negligible
effect. The corresponding distortion of the $\rho$ parameter can be
quantified by integrating out the heavy $A_X$ gauge boson: this leaves
a correction of order $(m_Z/\Lambda)^2$, which is safely small since
$\Lambda$ is experimentally constrained to be above a few TeV.

Bulk fermions in fundamental and symmetric representations allow
couplings to all the matter fermions, since their $U(1)^\prime$ charge
can be tuned to achieve the required hypercharge; the way in which the
hypercharge of the SM field is distributed as sum of $U(1)$ and
$U(1)^\prime$ charges is then completely fixed. In this model, one can
thus implement the construction described in the section 3,
without any additional complication. It turns out that four bulk
fermions $\Psi_a$ with $a=d,u,l,\nu$ (plus their mirrors $\tilde
\Psi_a$ with opposite parities) do the job, the quantum numbers of
bulk and brane fields with respect to $SU(3)_c\times SU(3)_w\times
U(1)^\prime$ and $SU(3)_c\times SU(2)_L\times U(1) \times U(1)^\prime$
being as follows:
\begin{eqnarray}
\begin{array}{lll}
\Psi_d: ({\bf 3},{\bf 3})_0^+ \;,\;\;\a
\mbox{couples to} \;\;\a
Q_L: ({\bf 3},{\bf 2})_{1/6,0} \;\,\mbox{and}\;\,
d_R: ({\bf 3},{\bf 1})_{-1/3,0} \;, \raisebox{15pt}{}\smallskip \\
\Psi_u: ({\bf \bar 3},{\bf 6})_0^- \;,\;\;\a
\mbox{couples to} \;\;\a
Q_R^c: (\bar {\bf 3},{\bf 2})_{-1/6,0} \;\,\mbox{and}\;\,
u_L^c: (\bar {\bf 3},{\bf 1})_{-2/3,0} \;, \smallskip \\
\Psi_l: ({\bf 1},{\bf 3})_{-2/3}^+ \;,\;\;\a
\mbox{couples to} \;\;\a
L_L: ({\bf 1},{\bf 2})_{1/6,-2/3} \;\,\mbox{and}\;\,
l_R: ({\bf 1},{\bf 1})_{-1/3,-2/3} \;, \smallskip \\
\Psi_\nu: ({\bf 1},{\bf 6})_{2/3}^- \;,\;\;\a
\mbox{couples to} \;\;\a
L_R^c: ({\bf 1},{\bf 2})_{-1/6,2/3} \;\,\mbox{and}\;\,
\nu_L^c:({\bf 1},{\bf 1})_{-2/3,2/3} \;. \medskip
\end{array}
\label{bulk4}
\end{eqnarray}
To achieve the most general flavour structure, we introduce three
generations $I=1,2,3$ of the above bulk fields. These can have
arbitrary kinetic matrices $(M_a)_{IJ}$ in flavour space, possibly
different for $a=d,u,l,\nu$. The couplings of the bulk fermions to the
three generations of SM boundary fields can involve generic matrices
$(\epsilon_i^a)_{IJ}$ in flavour space.  However, these can be made
proportional to the identity through a rotation of the bulk fields,
whose only additional consequence will be to change the kinetic
matrices $(M_a)_{IJ}$. Without loss of generality, we can therefore
set the couplings to flavour-blind constants, parametrized by
dimensionless coefficients $\epsilon_1^a$ and $\epsilon_2^a$ for
left-handed and right-handed fields.

\subsection{Mass matrices}

The computation of the induced masses for the SM matter fields
proceeds exactly as in section 3, the only novelty being the
non-trivial flavour structure of the bulk masses. The latter can be
written as $M_a = E_a^{\dagger} M_a^D F_a$, where $E_a$, $F_a$ are
unitary matrices and $M_a^D$ is diagonal, and the kinetic term of the
bulk fields can thus be diagonalized in flavour space by redefining
$\Psi_a^\prime = E_a \Psi_a$ and $\tilde \Psi_a^\prime = F_a \tilde
\Psi_a$.  By doing so, one gets a diagonal parameter $\lambda^a = \pi
R M_a^D$ for bulk fields, but the couplings between boundary fields
and the bulk fields $\Psi_a^\prime$ become non-diagonal and involve the
matrices $E_a$. The right-handed part of the boundary fields $a$,
which couple only to the corresponding bulk fermion $\Psi_a^\prime$,
can be diagonalized by redefining $a_R^\prime = U_a a_R$ with $U_{d,l} =
E_{d,l}$ and $U_{u,\nu} = E_{u,\nu}^*$.  For the left-handed fields,
instead, this cannot be done, since each of them couples to two
different bulk fields; this will be a first source of non-trivial
mixing in the mass matrices. At this point, all the couplings are
diagonal, except those mixing the boundary fields $Q_L$, $L_L$,
$Q_R^{c}$ and $L_R^{c}$ to the bulk fields $\Psi_{d}^\prime$,
$\Psi_{l}^\prime$, $\Psi_{u}^\prime$ and $\Psi_{\nu}^\prime$, which
involve the matrices $U_d$, $U_l$, $U_u$ and $U_\nu$.

The presence of the matrices $U_a$ affects the results of section 3 in
the following way. The contribution to the wave function $Z_1^a$ of
the left-handed field $a_L$ from the bulk field $\Psi_b^{\prime}$, call
it $Z_1^a(\Psi_b^{\prime})$, is changed to $\tilde
Z_1^{a}(\Psi_b^{\prime}) = U_b^\dagger Z_1^a (\Psi_b^{\prime}) U_b$, and
the new total wave function $\tilde Z_1^{a} = \sum_b \tilde
Z_1^{a}(\Psi_b^{\prime})$ is not diagonal.  The wave-function
corrections $Z_2^a$ for the right-handed fields $a_R$ are instead
unchanged and diagonal: $\tilde Z_2^a = Z_2^a$. Finally, the mass
$m^a$ induced for the boundary fields is changed to $\tilde m^a =
m^a U_a$. In order to determine the new physical mass, one has to
diagonalize the kinetic term of the left-handed fields. This is
achieved through a unitary transformation; writing $\tilde Z_1^a = V_a
Z_1^{D a} V_a^\dagger$, with $Z_1^{D a}$ diagonal, one can redefine
the left-handed fields as $a_L^\prime = V_a^\dagger a_L$, to obtain a
kinetic term for boundary fields which is diagonal in flavour space.  The
wave functions are now all diagonal and given by $Z_1^{D a}$ and
$Z_2^a$, whereas the mass matrices are all non-diagonal and given by
$m^a U_a V_a$.  Rescaling finally the wave-function factors, one finds
the following physical mass matrices:
\begin{equation}
(m_{\rm phys}^a)_{IJ} = \frac {m^a_{II}(V_a U_a)_{IJ}}
{\sqrt{(Z_2^{a})_{II} (Z_1^{Da})_{JJ}}}
\ \ ({\rm no \; sum \; on \;} I,J)\;.
\end{equation}

\subsection{Anomalies}

We now briefly comment on the issue of anomalies, paying attention to
their distribution over the internal space. Since the bulk fermions
are strictly vector-like, the only anomalies that can arise come from
the SM fermions living at the fixed-points, and depend on how these
are distributed among the two different fixed-points.

In the case where all SM fermions are located at the same fixed-point,
all anomalies that do not involve the extra $U(1)_X$ gauge field
vanish, thanks to the usual cancellations arising for the SM spectrum
of fermions.  We are then left with localized mixed anomalies
involving the $U(1)_X$ gauge field, which can be cured by means of a
4D version of the Green--Schwarz mechanism (GS) \cite{GS}. One
introduces a neutral 4D axion at $y=0$, transforming non-homogeneously
under the $U(1)_X$ symmetry, with non-invariant 4D Wess--Zumino
couplings compensating for the one-loop anomaly. In this way all mixed
$SU(3)_c\times SU(2)_L \times U(1)_Y \times U(1)_X$ gauge and
gravitational anomalies are cancelled and the axion is eaten by the
$U(1)_X$ gauge boson, which becomes massive and decouples.

For other distributions of the SM fermions such that all the SM
anomalies are still cancelled locally, a GS mechanism is again sufficient
to cancel all remaining anomalies involving the $U(1)_X$ gauge field.
However, two neutral axions are now needed, one at $y=0$ and one at
$y=\pi R$, with non-invariant 4D Wess--Zumino couplings. One
combination of axions is again eaten by the $U(1)_X$ gauge boson, but
the other combination remains as a physical massless axion in the
low-energy spectrum.

For a completely generic distribution of matter, for which not even
the SM anomalies are locally cancelled, the situation is more
complicated.  In order to locally cancel the SM anomalies, one has to
introduce a bulk Chern--Simons term with jumping coefficient
\cite{anomaly}-\cite{CSnonab}, which can be naturally generated by
integrating out certain massive states \cite{sst}. Since the
hypercharge is embedded into a non-Abelian group in the bulk, this is
however not sufficient to let all of the anomalies flow on a single
fixed-point. One then has to introduce also two neutral axions, one at
each fixed-point, to locally cancel all the remaining anomalies
involving $U(1)_X$. As before, the $U(1)_X$ gauge boson gets a mass,
but a combination of axions remains massless.

\subsection{Quantitative analysis}

This simple 5D orbifold model we constructed has all the qualitative
features to represent a possible interesting extension of the SM,
where the electroweak scale is stabilized without supersymmetry
and the hierarchy of fermion masses is explained by the non-local
origin of the Yukawa couplings.  As mentioned at the end of section 4,
the quantities $m_H$, $m_W$ and $m_Z$, and especially their ratios,
can be reliably computed, having a mild dependence on the cut-off
$\Lambda$. In order to get a better understanding of the range of
validity of our model as an effective field theory, it is however
necessary to know the magnitude of $\Lambda$. A reasonable estimate
is obtained by defining $\Lambda$ as the energy scale where
the basic higher-dimensional gauge interactions become strong. NDA
then yields $\Lambda \sim (l_D/g_D^2)^{1/(D-4)}$, where $g_D$ is the
higher-dimensional gauge coupling and $l_D = (4\pi)^{D/2} \Gamma(D/2)$
in $D$ space-time dimensions. The predictive power of the effective
theory at the compactification scale $M_{\rm c}$ is therefore governed
by the 4D effective coupling $g_4 = g_D M_{\rm c}^{(D-4)/2}$, since the
small parameter controlling corrections due to non-renormalizable operators
is given by $M_{\rm c}/\Lambda \sim (g_4^2/l_D)^{1/(D-4)}$, and can
thus be lowered by effects that tend to increase the effective gauge
coupling\footnote{Of course $\Lambda$ is always larger than $M_c$,
  since at lower energies the theory returns 4D, and the above
  estimate for $\Lambda$ is accurate only as long as $\Lambda \gg
  M_{\rm c}$.}.

In our 5D model, the loop factor is given by $l_5 = 24 \pi^3$ and the
5D and 4D gauge couplings are related by $g_4 = g_5/\sqrt{2 \pi R}$,
so that $\Lambda \sim (12 \pi^2)/g_4^2$. Considering respectively the
strong and the weak interactions, this would give roughly $\Lambda_c
\sim 10/R$ and $\Lambda_w \sim 100/R$. This means that $\Lambda$ can
be identified with $\Lambda_c$ and the theory is indeed reasonably
predictive. In particular, the universal wave-function corrections for
the $H$, $Z$ and $W$ fields are proportional to $(g_{5,w}^2/l_5)
\Lambda$, that is $\Lambda_c/\Lambda_w$, and therefore represent small
corrections that can be neglected.  We can then go a step further and
ask whether this minimal 5D model is also quantitatively a
phenomenologically viable model. As we will see, this not quite the
case, because the values predicted by the model for $1/R$, $m_H$ and
$m_{\rm top}$ are too low.

The crucial parameter that sets the scale of the model is $\alpha$,
whose value is determined by minimizing the full effective potential
$V(\alpha)$, as in section 4. Generically, the most relevant fermionic
contribution to the potential is induced by the bulk fermion in the
symmetric of $SU(3)_w$ that gives mass to the top quark, as expected.
Neglecting the effect of the other bulk and boundary fermions, we have
numerically analysed the form of $V(\alpha)$ as a function of
$\lambda^{\rm top}$ and of the bulk-to-boundary couplings
$\epsilon_i^{\rm top}$. As discussed in the previous section, the
lowest non-trivial value for $\alpha$ that we get is $\alpha \simeq
0.16$ for $\delta = 0$ and $\alpha \simeq 0.12$ for $\delta = 1$ (see
fig.~\ref{fig:veff1}), which by means of (\ref{MW}) implies $1/R \sim
500$ GeV. This is in conflict with experimental bounds for models such
as ours, with localized interactions that do not conserve KK momentum,
which require roughly $1/R \sim$ few TeV \cite{Delgado:1999sv}.  The
Higgs mass computed using (\ref{MH}) is also too low, at most $m_H
\sim 30$ GeV. Finally, the top Yukawa coupling arising from
(\ref{mu})--(\ref{Zid}) turns out to be too small for any value of
$\epsilon_i^{\rm top}$ and $\lambda^{\rm top}$, giving a bound
$m_{\rm top}\leq 65$ GeV. We actually have also evaluated the top mass by
numerically solving the exact transcendental equation arising from the
effective potential, as discussed in section 4, but the deviations
from the approximate relations (\ref{mu})--(\ref{Zid}) turn out to be
very small and negligible.

It is interesting to notice that all these problems could be
alleviated if $V(\alpha)$ had minima at lower $\alpha$. For $\alpha
\sim 0.01$, for instance, $1/R$ would be well above the current
experimental bounds and $m_H$ would increase up to more than $100$
GeV. The top mass slightly increases, but still
$m_{\rm top}\leq 110$ GeV.  This is not yet completely satisfactory,
but it goes in the right direction. One should also remember that
these predictions, in particular the top mass, could be affected by
large corrections if the cut-off of the model is low.  One should also
remind that all our analysis is based on minimizing the one-loop
effective potential $V(\alpha)$, and it is not easy to estimate how
much higher-loop contributions to $V(\alpha)$ can alter the actual
value of $\alpha$. The latter, as we have seen, is the crucial parameter
in these models, and it would be extremely interesting to find a
mechanism that gives rise to a potential with such low values of
$\alpha$ without spoiling the nice features of the model.

\subsection{Possible extensions}

\begin{figure}[t]
  \begin{center}
    \begin{tabular}{c c}
      \hspace{-10pt}
      \includegraphics[width=0.5\textwidth]{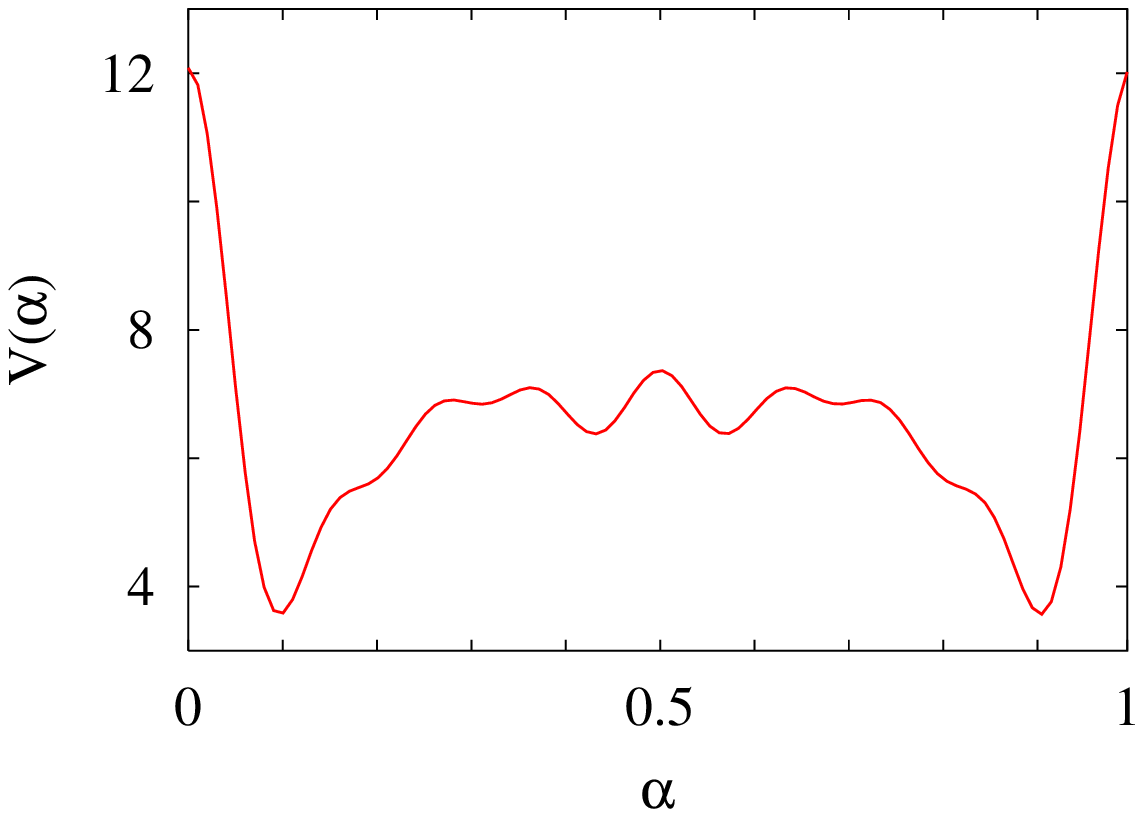} &
      \hspace{-10pt}
      \includegraphics[width=0.5\textwidth]{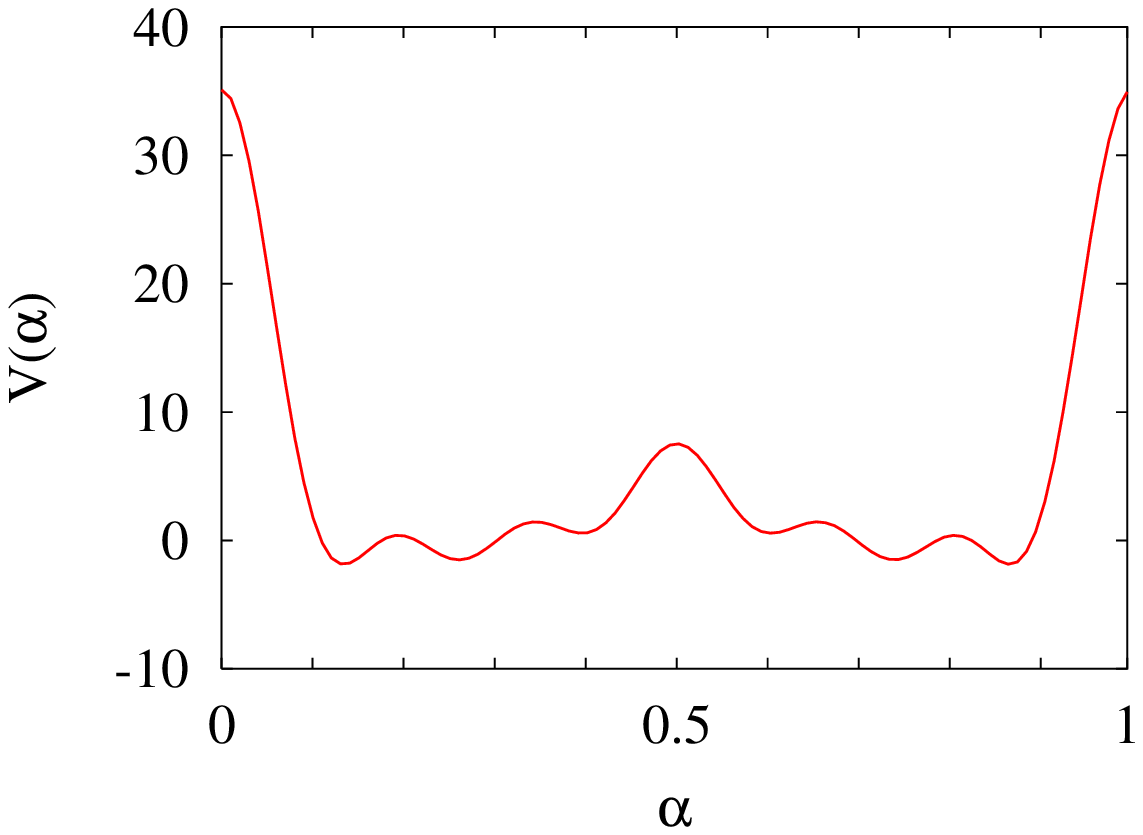} \\
    \end{tabular}
  \end{center}
  \caption{\small
      The full effective potential (in arbitrary units) in the
      presence of high-rank bulk fermions. Left: $r=8$, $\lambda=3.47$,
      $\epsilon_1=\epsilon_2=9$ and $\delta=0$, resulting in $m_H=112$
      GeV and $1/R=830$ GeV. Right: $r=6$, $\lambda=2.23$,
      $\epsilon_1=7$, $\epsilon_2=1$ and $\delta=1$, resulting in $m_H=104$
      GeV and $1/R=600$ GeV.}
  \label{fig:veffr}
\end{figure}

As anticipated in the introduction, a possible way to lower $\alpha$
is to consider 5D massive bulk fermions in large $SU(3)_w$
representations.  For example one can take a completely symmetric
representation of large rank $r$, with dimension $d(r) =
(r+1)(r+2)/2$. This contains components of charge $q=k$, where $k$
is an integer ranging from $0$ to $r$. The multiplicities of the charged
states with $k \neq 0$ are found to be $N_k = 1 + [(r-k)/2]$, where
$[\dots]$ denotes the integer part. This information allows us to
compute the contribution of this field to the effective potential by
summing up the contributions of all the charged components computed
with eq.~(\ref{Va-xCorr}). We find that if the rank $r \gg 1$ and the
parameter $\lambda$ controlling the mass is not too large, we can have
minima for $\alpha \sim 1/r$, and furthermore the second derivative
$V^{\prime\prime}$ at the minimum grows very quickly with $r$, leading
indeed to a substantial improvement of the situation. With $r = 6$,
$\lambda\sim 2.2$ and $\delta = 1$, we obtain $\alpha \sim 0.13$,
corresponding to $1/R\sim 0.6$ TeV and $m_H \sim 104$ GeV, while for
$r = 8$, $\lambda\sim 3.5$ and $\delta = 0$, we get $\alpha \sim 0.096$,
corresponding to $1/R\sim 0.8$ TeV and $m_H \sim 112$ GeV (see
fig.~\ref{fig:veffr}).

It should however be noted that matter fields in large
representations of the gauge group will induce electroweak quantum
corrections that are enhanced by large group-theoretical factors
$T(r)$.  The scale at which the weak coupling becomes non-perturbative
is therefore lowered: $\Lambda_w \rightarrow \Lambda_w /T(r)$. It is
difficult to give a precise quantitative estimate of $T(r)$, because
it is not universal.  To get an order of magnitude, one can use the
Dynkin index of the representation, which in our case is found to be
$T(r)=r(r+1)(r+2)(r+3)/48$ \cite{slansky}. This shows that $\Lambda_w$
rapidly decreases as $r$ increases too much. When $\Lambda_w$ becomes
comparable with $\Lambda_c$, the wave-function corrections to the
physical masses get out of control, and only ratios of these masses
can be predicted, as long as $\Lambda_w$ does not get too close to $1/R$.
We believe that values up to $r \sim 10$ could be reasonable.

\section{Localized gauge kinetic terms}

In the orbifold models we consider, gauge fields have certainly a bulk
kinetic term, but no symmetry forbids the occurrence of additional
localized kinetic terms. It is therefore interesting to consider the
general case in which all of these are present with arbitrary
coefficients, and study the consequences on model building. This kind
of situation was first considered in \cite{Dvali:2000rx}, in the
context of non-compact higher-dimensional theories, and more recently
in \cite{Carena:2002me} (see also \cite{delAguila:2003bh}) for compact
orbifolds\footnote{Localized gauge kinetic terms do also naturally
  occur at tree level in certain string theory models; see e.g.
  \cite{abd}.}.  We will consider for simplicity $H$-universal
localized kinetic terms for 4D gauge fields only, described by two
couplings $l_{1,2}$ with mass dimension $-1$. Localized terms
involving $A_5$ are allowed, but their presence considerably complicates
the analysis, and we therefore discard them. Denoting 5D and
4D indices with $M,N$ and $\mu,\nu$ respectively, the Lagrangian for
the $SU(3)_w$ gauge fields is then given by
\begin{equation}
\label{Lagbos}
\mathcal{L}_g = -\frac 12 {\rm Tr}\, F_{MN}F^{MN}
-  \Big(l_1\,\delta(y) + l_2\,\delta(y-\pi R) \Big)
{\rm Tr}\, F_{\mu \nu} F^{\mu \nu} \;.
\end{equation}
As usual, it will be convenient to introduce dimensionless parameters
relating the coefficients of the localized kinetic terms to the length
of the orbifold: $c_i = (\pi R)^{-1}l_i$. For simplicity, we do not
add localized terms associated with the $U(1)^\prime$ gauge field. It
actually turns out that their main effect would be a simple rescaling
of the coupling constant $g^\prime$ and so we do not lose in
generality by discarding them.

\subsection{Spectrum}

The wave functions and KK spectrum for the 4D gauge fields
$A_\mu^{(i)}$ are distorted by the localized couplings appearing in
(\ref{Lagbos}).  The effect of the non-vanishing VEV $\alpha$ is most
conveniently taken into account by adopting the point of view of
twisted boundary conditions, in which each component with definite
charge $q$ (in a diagonal basis) satisfies $f_n(y+2\pi R, q\alpha) =
e^{2i \pi q \alpha} f_n(y, q\alpha)$.  The differential equation
defining the wave functions $f_n(y, q\alpha)$ of the mode with mass
$m_n(q\alpha)$ is found by proceeding as in \cite{Carena:2002me}. It
reads:
\begin{equation}
\Big\{\partial_y^2 + m_n^2(q \alpha) \Big[1 + 2 \pi R\,c_1\, \delta(y)
+ 2 \pi R\,c_2\, \delta(y-\pi R)\Big]\Big\} f_n(y;q \alpha) = 0 \;.
\label{fneq}
\end{equation}
The general solution of this equation in the interval $[-\pi R,\pi R]$
has the form
\begin{eqnarray}
f_n(y;q \alpha) = {\cal N}_n(q \alpha) \left\{
\begin{array}{l}
\cos (m_n y) + \beta_n^- \sin (m_n y)
\;,\;\;\mbox{$y \in [-\pi R,0]$} \medskip \\
\cos (m_n y) - \beta_n^+ \sin (m_n y)
\;,\;\;\mbox{$y\in [0,\pi R]$}
\end{array}
\right. \;.
\label{fnsol}
\end{eqnarray}
The constant ${\cal N}_n$ is a normalization factor defined in such a
way that $\int_{0}^{2\pi R} |f_n|^2 dy = 1$. The parameters
$\beta_n^\pm$ are fixed by the twisted boundary conditions and the
discontinuity at $y=0$, and read:
\begin{equation}
\beta_n^{\pm} = e^{\pm i \pi q \alpha} \sec (\pi q \alpha) (\pi R m_n) c_1
\mp i \tan(\pi q \alpha) \cot(\pi R m_n) \;.
\end{equation}
Finally, the spectrum is determined by the discontinuity at $y = \pi
R$, which enforces the following transcendental equation for the
dimensionless eigenvalues $\xi_n = \pi R m_n$:
\begin{equation}
2(1 - c_1 c_2\, \xi_n^2) \sin^2 \xi_n
+ (c_1 + c_2)\, \xi_n \sin 2\xi_n
- 2\,\sin^2 (\pi q \alpha) = 0\,.
\label{spectrumgauge}
\end{equation}
For $\alpha=0$, eqs.~(\ref{fnsol})--(\ref{spectrumgauge}) reduce to
the equations derived in \cite{Carena:2002me}. For $c_i = 0$, they
also correctly reduce to the case of twisted gauge bosons; in
particular, the solution of eq.~(\ref{spectrumgauge}) is then $\xi_n =
\pi (n+q\alpha)$.

The deformation induced on the mass spectrum is one of the most
important effects of localized kinetic terms. The mass of the lightest
modes can be determined by solving the transcendental equation
(\ref{spectrumgauge}) in the limit $\xi_0 \ll 1$. One finds in this
way $\xi_0 \sim \sin (\pi q \alpha)/\sqrt{1+c_1+c_2}$, and
self-consistency of the assumption $\xi_0 \ll 1$ requires that $\alpha
\ll 1$ and/or $c_i \gg 1$. This expression must be compared with the
value $\xi_0 = \pi q \alpha$ of the standard case $c_i=0$, and shows
that the relation between masses of light gauge boson modes with
different charges $q$ is distorted if $c_i\neq 0$. The masses of heavy
KK modes ($n \ge 1$) are deformed as well, and tend to become lighter;
one finds that $\xi_n \rightarrow (n-1)$ if $c_1 \sim c_2$ whereas
$\xi_n \rightarrow (n-1/2)$ if $c_1 \gg c_2$ or vice versa.

The deformation induced on the wave functions of the KK modes is also
particularly relevant, because it affects the concept of 4D effective
gauge coupling constant, which is no longer universal. More precisely,
one can define an effective gauge coupling $g_{4,q,n}(y)$ between
matter fields localized at $y=0,\pi R$ and the $n$-th KK mode of the
gauge bosons with charge $q$; defining the quantity $Z_n(q\alpha) = 1
+ 2\pi R\, c_1 |f_n(0;q\alpha)|^2 + 2\pi R\, c_2 |f_n(\pi
R;q\alpha)|^2$, this is found to be
\begin{equation}
g_{4,q,n}(y) = g_5 \frac{|f_n(y;q\alpha)|}
{\sqrt{Z_n(q\alpha)}} \;.
\label{geffective1}
\end{equation}
Similarly, one can define a coupling $g_{4,q_i,n_i}$ between three
gauge bosons with KK modes $n_i$ and charges $q_i$, which is given
by\footnote{In (\ref{geffective2}), the integral over $y$ should be
defined in the interval $[-\epsilon,2 \pi R-\epsilon]$ to correctly normalize
the boundary contributions.}
\begin{equation}
g_{4,q_i,n_i} = g_5 \int_0^{2\pi R} \!\!\!\! dy
\Big[1+ 2\pi R\,c_1\,\delta(y) + 2\pi R\,c_2\,\delta(y-\pi R)\Big]
\prod_{i=1}^3 \frac {|f_{n_i}(y;q_i\alpha)|}
{\sqrt{Z_{n_i}(q_i\alpha)}} \;.
\label{geffective2}
\end{equation}
The above equations describe an important distortion of the gauge
coupling.  In particular, the strength of the coupling depends on the
type of modes and their location. Contrary to what happens for
$\alpha=0$ \cite{Carena:2002me}, eqs.~(\ref{geffective1}) and
(\ref{geffective2}) represent a distortion also for the zero-mode
couplings, because the wave function $f_0$ becomes non-constant and
$q$-dependent for $\alpha \neq 0$.

\subsection{Contribution to the effective potential}

The contributions of gauge fields to the Higgs effective
potential are modified by the localized kinetic terms as well, and
must be recomputed.  It is convenient to use the background
gauge-fixing condition $\bar D_M A^M = \partial_M A^M + i g_5 [\bar
A_M, A^M] = 0$, where $\bar A_M = \delta_{M5}(\alpha/g_5 R) \lambda^7$
is the background field.  This gauge-fixing is not affected by the
localized boundary terms and the ghost fields $\eta$ therefore have
only a bulk kinetic term.  After gauge-fixing, the bulk and ghost
kinetic operators are diagonal and proportional to $\bar D^P \bar
D_P$, but the boundary kinetic operator involves the transverse
projector $\bar D^\rho \bar D_\rho \eta^{\mu\nu} - \bar D^\mu \bar
D^\nu$. As usual, a change of basis is required to diagonalize the
couplings to $A_5$, and there are two modes with charge $q=1$ and
one mode with charge $q=2$, both for ghost and gauge fields.
After KK decomposition, the kinetic operator of the ghosts and
the internal component of the gauge fields have a simple diagonal
form, whereas the one of the four-dimensional components of the
gauge fields is deformed in a non-trivial way. They are given by
\begin{eqnarray}
K^{\rm A_5,gh}_{mn} \a=\a \delta_{m,n}(p^2 + m_{n}^2) \;,\\
K^{A_\mu}_{mn,\mu\nu} \a=\a \delta_{m,n} \eta_{\mu \nu} (p^2 + m_{n}^2)
+ (c_1 + (-1)^{m+n} c_2) (\eta_{\mu \nu} p^2 - p_\mu p_\nu)\;.
\end{eqnarray}
The contribution to the effective action of each type of mode with
fixed charge $q$ is given by $\Gamma(q\alpha) = 1/2\, {\rm ln}\, {\rm
  det} [K^{A_\mu} K^{A_5}(K^{\rm gh})^{-2}(q\alpha)]$. The determinant
over the KK and vector indices can be explicitly performed. For
$K^{A_5}$ and $K^{\rm gh}$ this is trivial; for $K^{A_\mu}$, the KK
part can be done by considering a finite-dimensional truncation and
recursively increasing the dimensionality, and the vector part just
produces a factor of $3$ in the exponent coming from the trace of the
transverse projector. The result is finally
\begin{equation}
{\rm Det}\Bigg[\frac {K^{A_M}}{(K^{\rm gh})^{2}}\Bigg]
= \prod_n (p^2+m_{n}^2)^3
\Bigg[\prod_{i=1}^2 \Big(1 + c_i \sum_n \frac{p^2}{p^2+m_{n}^2}\Big)
- \prod_{i=1}^2 \Big(c_i \sum_{n} \frac {p^2 (-1)^{n}}{p^2+m_{n}^2}\Big)
\Bigg]^3\,.
\end{equation}
As for the fermions, the contribution of each mode to the potential
naturally splits into the standard bulk part and a boundary part
encoding the effects of the localized interactions. The bulk part
$V_g^A$ is given by eq.~(\ref{V5gauge1}). The boundary corrections can
instead be written as
\begin{eqnarray}
V^{c_i}_{g}(q \alpha) \a=\a \frac {3}{16\pi^6 R^4}
\int_0^\infty \!\!\! dx\,x^3\,{\rm ln}
\Bigg[\prod_{i=1}^2 {\rm Re} \Big[1 \!+\! c_i x f_0(x,q\alpha) \Big]
- \prod_{i=1}^2 {\rm Re} \Big[c_i x f_1(x,q\alpha) \Big] \Bigg] \;.
\hspace{20pt}
\label{V5gauge2}
\end{eqnarray}

\begin{figure}[t]
  \begin{center}
      \hspace{-10pt}
      \includegraphics[width=0.6\textwidth]{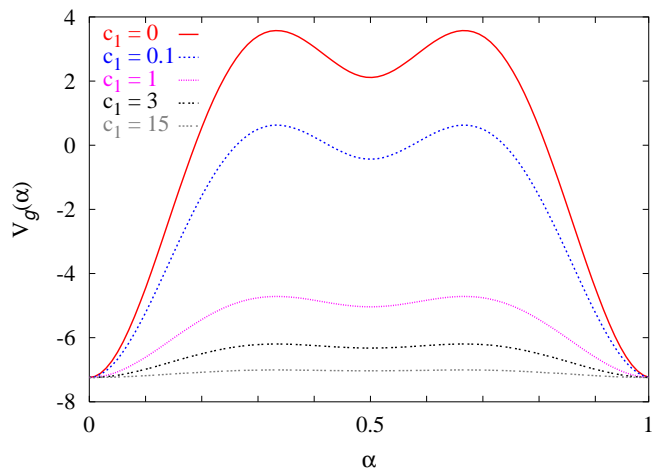}
  \end{center}
  \caption{\small
      Gauge contribution to the effective potential (in
      arbitrary units) in the presence of localized gauge kinetic
      terms, with $c_{2}=0$ and
      increasing values of $c_1$. }
  \label{fig:vgeff}
\end{figure}

The total contribution of gauge fields to the one-loop effective
potential is finally obtained by summing up the contributions of the
three charged modes; it is given by
\begin{equation}
V_g(\alpha) = 2 V^{A}_{g}(\alpha) + V^{A}_{g}(2\alpha)
+ 2 V^{c_i}_{g}(\alpha) + V^{c_i}_{g}(2\alpha) \,.
\label{Vgc}
\end{equation}
Again, the result is finite, thanks to the exponentially soft UV
behaviour of the functions $f_\delta$. Notice, moreover, that for $c_i
\gg 1$, the $\alpha$-dependence in the boundary contribution tends to
exactly cancel the $\alpha$-dependence in the bulk contribution (see
fig.~\ref{fig:vgeff}). This is most easily seen by putting
(\ref{V5gauge1}) and (\ref{V5gauge2}) together and simplifying the
argument of the logarithm, which becomes $\prod_i(\cosh x
+ c_i x \sinh x) - \cos^2 (\pi \alpha)$.  Note that this expression
is proportional to eq.~(\ref{spectrumgauge}) after the analytic
continuation $x \rightarrow i \xi_n$, and has therefore the same
zeros. This constitutes a non-trivial consistency check of our result
(\ref{Vgc}). Indeed, the latter could have been computed as the
effective action of the new eigenstates, which would have led to a
logarithm with an argument proportional to $\prod_n (x^2 + \xi_n^2)$,
with zeros at $x = i \xi_n$. In this approach, however, performing
the product over the KK modes is non-trivial (see for instance
\cite{Ponton:2001hq}). It is also interesting to observe that, as
in the case of the fermions, the boundary contribution to the
effective potential can be rewritten in terms of the determinant of a
two-by-two matrix encoding the fixed-point-to-fixed-point propagation.
The localized gauge kinetic terms (\ref{Lagbos}) do not break the shift
symmetry $A^a_5\to A^a_5 + \partial_5 \xi^a$, and thus no direct divergence
is expected in the gauge contribution to the potential at any order
in perturbation theory.

\subsection{Effects on model building}

The deformations of the mass spectrum, gauge coupling constants and
induced effective potential that we have described in the last two
subsections have important consequences on model building. The
analysis for generic values of $\alpha$ and $c_i$ is however quite
involved, and we therefore consider only the special situation in
which $c_1 \gg 1$ and $c_2 \ll 1$, for which a substantial
simplification occurs. In this limit, the zero-mode wave-function
$f_0$ reduces to a $c$-independent linear profile.
Equations~(\ref{geffective1}) and (\ref{geffective2}) then yield
$g_{4,q,0}(0) \simeq g_4$, $g_{4,q,0}(\pi R) \simeq g_4 \cos (\pi
q\alpha)$ and $g_{4,q,0} \simeq g_4$, with
\begin{equation}
g_4 = \frac {g_5}{\sqrt{c}\,\sqrt{2 \pi R}} \;.
\label{g4g5}
\end{equation}
For non-zero modes, a stronger suppression factor is found for the
couplings at $y=0$, whereas those at $y=\pi R$ remain finite in the
limit of large $c$.  As in the case of \cite{Carena:2002me}, this
phenomenon tends to suppress four-fermion operators induced from
boundary fields at $y=0$ by the exchange of heavy KK modes of the
gauge bosons, in spite of the fact that these are now lighter.
Correspondingly, the bounds on $R$ become milder. The $W$ and $Z$
gauge bosons, that arise from states with charge $1$ and $2$,
respectively, now have masses equal to\footnote{Here we are neglecting
  the mixing between the $Z$ and the anomalous $A_X$ boson that
  results in a further deformation in $m_Z$.}
\begin{equation}
m_W \simeq \frac {\sin (\pi \alpha)}{\pi \sqrt{c}\, R} \;,\;\;
m_Z \simeq \frac {\sin (2 \pi \alpha)}{2 \pi \sqrt{c}\, R}
\,\sec \theta_W  \;.
\label{mwmz}
\end{equation}
Finally, the expression for the Higgs mass is affected as well,
because the relation between the 5D and 4D couplings is modified
according to (\ref{g4g5}), and one finds:
\begin{equation}
m_H \simeq \frac{g_4 R}2 \sqrt{c} \, \sqrt{V^{\prime\prime}(\alpha)} \;.
\label{MHdist}
\end{equation}

We see that a sizable factor $c$ results in further improvements.  The
$W$ mass is lowered, so that the experimental bound on $R$ can be
satisfied with higher values of $\alpha$ and becomes therefore less
restrictive. Taking this the other way around, large values of $1/R$
are easier to achieve. The problem of obtaining a reasonable value for
$m_{\rm top}$ is basically solved for high enough values of $c$. As a
rough estimate, one would need $c\sim 100$ if both the left- and
right-handed components of the top field are at the same fixed-point,
and $c\sim 10$ if they are at different ones.  The Higgs mass
is also enhanced, and gets higher at fixed VEV $\alpha$.  All the
problems of our original 5D model, namely the too low values for
$1/R$, $m_H$ and $m_{top}$, can thus be solved by adding large
localized kinetic terms. However, even in the limit of very large $c$,
an unwanted distortion remains.  As can be seen from (\ref{mwmz}), the
$\rho$ parameter does not depend on $c$, and it is given by
\begin{equation}
\rho = \frac {m_W^2}{m_Z^2 \cos^2 \theta_W} \simeq \sec^2 (\pi \alpha) \neq 1 \;.
\end{equation}
Moreover, the effective gauge couplings of possible boundary fields
located at $\pi R$ are deformed by a factor $\cos(\pi \alpha)$ for
charged interactions mediated by the $W$ and $\cos(2 \pi \alpha)$ for
the neutral ones mediated by the $Z$. Once again, a phenomenologically
acceptable situation therefore seems to require very low values of
$\alpha$, which in turn is dynamically determined by the radiatively
generated Higgs effective potential.

As already mentioned, the contribution of the gauge fields to the
effective potential is also strongly deformed by localized kinetic
terms. For $c\gg 1$, its $\alpha$-dependence gets suppressed and the
total effective potential is dominated by the fermion contributions.
In this case, the good features that we found at the end of section 4,
namely the possibility of obtaining values for $\alpha$ lower than the
usual ones, thanks to the boundary couplings, is ruined and one typically
gets back values close to $\alpha \sim 0.3$. This clearly goes against
what is needed to exploit the good features associated with localized
gauge couplings. In particular, $\alpha \sim 0.3$ gives an
unacceptable value for $\rho$. Owing to the enhancement of the bulk
electroweak coupling, the scale where the latter becomes
non-perturbative is lowered, $\Lambda_w \rightarrow \Lambda_w/c$, as
for the case of high rank representations discussed in section 5.4.
This represents another limitation to an increasing of $c$, but
values up to $c \sim 15$ appear to be reasonable. Notice that for
$c\sim 15$ and $\alpha\sim 0.3$, one would get $1/R \sim 1$ TeV, which
turns out to be compatible with electroweak precision tests, thanks to
the fact that the couplings between matter and KK modes of the gauge
fields are suppressed.

Summarizing, we see that the presence of localized gauge kinetic terms
can drastically improve the situation, but only if they are accompanied
by low values of $\alpha$, which allow some control on the unwanted
deformations that these localized terms necessarily produce. Unfortunately,
these low values for $\alpha$ do not appear to be generated in minimal
situations. It would therefore be again of great help to have at our
disposal some mechanism that provides an additional contribution
to the potential that could lower the VEV $\alpha$, without distorting
the electroweak symmetry breaking. As already mentioned in section
5.4, one possibility consists in introducing extra bulk fermions in
large representations of $SU(3)_w$. Although
rather unusual, such an additional large-rank heavy fermion would lead
to an optimal situation when combined with localized gauge kinetic
terms.

It would be interesting to study what happens for generic values of
the $c_i$'s, because it is not excluded that all the deformations
induced by these terms could conspire, in particular situations, to
yield a phenomenologically viable model. On the other hand, it should
be recalled that localized terms, even if not introduced at tree level,
are radiatively generated in the theory and thus a proper study of their
effects is necessary to draw definite conclusions on model building
in this context.

Let us conclude this section by noting that the above considerations
are valid as long as one introduces localized gauge kinetic terms for
$A_\mu$ only. As already said, this is not a necessary restriction
and a localized term involving the 5D field strength $F_{MN}$ could be
considered. The analysis of this case is complicated by the presence of
derivatives along the internal directions and the computation of the
effective potential seems much more involved. The KK spectrum and wave
functions of the 5D gauge fields could be quite different, in particular
for the zero-mode sector, and it is not excluded that this case could
be phenomenologically interesting.

\section{Outlook}

We have studied in detail various aspects of orbifold models with
unification of gauge and Higgs fields, ordinary matter localized at
fixed-points, and additional heavy fermions in the bulk. We have also
analysed the effect of having large localized gauge kinetic terms in
these models. Electroweak symmetry breaking occurs at the quantum
level through a rank-reducing Wilson-line symmetry breaking and is
transmitted to matter at the boundaries by the massive bulk
fermions. The main advantage of this mechanism is that the flavour
structure of the SM can be achieved in an elegant way, without spoiling
the stability of the Higgs potential.

We have presented a simple prototype example in 5D, based on the above
structure and the gauge group $SU(3)_c\times SU(3)_w \times U(1)^\prime$.
For its minimal version, we find that $1/R$, $m_H$ and $m_{\rm top}$
turn out to be too low, but acceptable values can be obtained, with a
moderate tuning of the parameters, by adding extra heavy bulk fermions and/or
localized gauge kinetic terms. By doing so, however, the predictive power
of the model is lowered. Most importantly, we have seen that in the presence
of localized gauge kinetic terms the electroweak sector of the theory is
distorted in a non-universal and unwanted way.

At this stage, the prototype 5D models that we presented cannot be
considered neither as viable nor as ruled out.  To this purpose, we
think that a more careful phenomenological analysis in needed, which
should take systematically into account the effect of localized kinetic
terms for bulk fields and possible extra massive bulk fermions.
On the other hand, the general structure that we have illustrated
can be applied to similar constructions in more than five dimensions
as well. The main new feature is the presence of a tree-level quartic
potential for the Higgs fields arising from the decomposition of the
higher-dimensional gauge kinetic term. The electroweak symmetry breaking
still occurs radiatively, but the presence of the tree-level term can help
in achieving a larger Higgs mass. In particular, 6D models represent
the minimal version of this possibility, with two Higgs doublets
\cite{Antoniadis:2001cv,Csaki:2002ur}. We plan to extend our analysis
to this kind of models in a future work.

\section*{Acknowledgements}

We would like to thank G.~Giudice, B.~Mele, R.~Rattazzi, A.~Romanino,
M.~Salvatori, A.~Strumia and F.~Zwirner for many fruitful discussions
and suggestions. We also thank C.~Cs\'aki, C.~Grojean and H.~Murayama
for e-mail correspondence. This research work was partly supported by
the European Community through a Marie Curie fellowship and the RTN
Program ``Across the Energy Frontier'', contract HPRN-CT-2000-00148.
L.S. acknowledges CERN, and C.A.S. and M.S. the University of Rome
``La Sapienza'', where part of this work was done.

\appendix

\section{Mode decomposition}

The mode decomposition of fields in various representations of the
gauge group $G$ in the presence of a projection $P$ and twist $T$ can be
easily obtained as follows. If we denote by $\hat \Psi_{\cal R}(y)$ a
field multiplet transforming in the representation ${\cal R}$ of $G$,
we have
\begin{equation}
\hat \Psi_{\cal R}(-y) = \eta_\Psi\,{\cal R}(P)\,\hat \Psi_{\cal R}(y) \;,\;\;
\hat \Psi_{\cal R}(y+2\pi R) = {\cal R}(T) \hat \Psi_{\cal R}(y)\;,
\label{modes}
\end{equation}
where ${\cal R}(P)$ and ${\cal R}(T)$ denote respectively the
embedding of the projection and twist in the gauge group in the
corresponding representation, and $\eta_\Psi=\pm 1$. In a basis in
which $P$ is diagonal, the first relation in (\ref{modes}) is
easily solved in terms of single-valued fields $\Psi_{\cal R}$. One
simply gets an expansion in cosines or sines for the various components,
according to the eigenvalue of the projection matrix $P$.  The second
relation in (\ref{modes}) is then satisfied by taking
\begin{equation}
\hat \Psi_{\cal R}(y) = {\cal R}[\Omega(y)] \, \Psi_{\cal R}(y)\;.
\label{psihatpsi}
\end{equation}
In eq.~(\ref{psihatpsi}), $\Omega(y)=\exp(i \alpha_a \tau^a y/R)$ when
$T$ is expressed as $T=\exp(2i\pi \alpha_a \tau^a)$, with $\tau^a$ the
generators of the Lie algebra of the group $G$. The field
$\hat \Psi_{\cal R}(y)$
automatically solves also the first relation in (\ref{modes}) because
${\cal R}[\Omega(-y)]={\cal R}[\Omega^{-1}(y)]$ and $T P T = P$ by
consistency.  Since (\ref{psihatpsi}) is simply a non-single valued
gauge transformation, we can alternatively work with the untwisted
fields $\Psi_{\cal R}$ only. In this gauge, the effect of the twist is
encoded in the VEV for $A_5$ induced by the gauge transformation:
$A_5=(-i/g) \Omega^\dagger(y)\partial_5 \Omega(y) = \alpha_a \tau^a/(g
R)$. In this case, the problem of finding the mode decomposition of
the field $\Psi_{\cal R}$ is reduced to the choice of a basis in which
its coupling with $\langle A_5\rangle$ is diagonal.

In the following, we will adopt this second point of view and consider
the decomposition of the untwisted fields $\Psi_{\cal R}$, for the case in
which ${\cal R}$ is the fundamental, symmetric and adjoint
representation of $G=SU(3)$, with $P$ and $T$ taken as in
(\ref{Rtwist}) and (\ref{Ttwist}).  It will be convenient to introduce
the factor $\eta_n$ defined to be $1$ for $n=0$ and $1/\sqrt{2}$ for
$n\neq 0$, as well as basic wave functions of even and odd modes:
\begin{equation}
f_n^+(y) = \frac 1{\sqrt{2 \pi R}} \cos \frac {ny}R \;,\;\;
f_n^-(y) = \frac 1{\sqrt{2 \pi R}} \sin \frac {ny}R \;.
\end{equation}
We will mainly focus on matter fermions, but our results easily generalize
to other fields. We will denote by $\Psi_\pm$ the left- and right-handed
components, which satisfy the first equation in (\ref{modes}) with
$\eta_\Psi=\pm$ respectively.

\subsection{Fundamental}

For the fundamental representation we have simply ${\cal R}(P)=P$ and
${\cal R}(T)=T$ in (\ref{modes}). It is convenient for later purposes to
express $\Psi_\pm$ as a sum over all integer modes, both positive and
negative; this is done by defining the negative modes of a given component
as the reflection of the positive modes: $\Psi_{-n} = \pm \Psi_n$,
depending on the parity of the component. In this way, the mode
expansion for the untwisted fields $\Psi_\pm$ is given by
\begin{equation}
\Psi_\pm(y) = \sum_{n=-\infty}^\infty \eta_n
\left(\matrix{
f_n^\mp(y)\, \psi_{n}^{\pm u} \cr
f_n^\mp(y)\, \psi_{n}^{\pm d} \cr
\pm f_n^\pm(y)\,\chi_{n}^{\pm} \cr}\;
\right)\;,
\label{dousingF}
\end{equation}
where we denoted by $\psi^u$, $\psi^d$ the up and down components of the
$SU(2)\times U(1)$ doublet, and by $\chi$ the $SU(2)\times U(1)$ singlet.

The basis in which the coupling of the three components of the triplet to
the VEV of $A_5$ is diagonal is reached by defining the following new fields:
\begin{equation}
\Psi_{n}^{\pm(1)} = \psi_{n}^{\pm u} \;,\;\;
\Psi_{n}^{\pm(2)} = \eta_n \Big(\psi_{n}^{\pm d} + \chi_{n}^{\pm}\Big) \;.
\label{fundredef}
\end{equation}
Notice that all the modes of $\Psi_n^{\pm(2)}$ are physical and
correspond to orthogonal combinations of the physical modes
$\psi_{n}^{\pm d}$ and $\chi_n^\pm$. In this new basis, the wave
function is rewritten as
\begin{equation}
\Psi_\pm(y) = \sum_{n=-\infty}^\infty
\left(\matrix{
\eta_n f_n^\mp(y)\, \Psi_{n}^{\pm(1)} \cr
f_n^\mp(y)\, \Psi_{n}^{\pm(2)} \cr
\pm f_n^\pm(y)\, \Psi_{n}^{\pm(2)} \cr}\;
\right)\;.
\label{wavefun}
\end{equation}
The action of the twist is now diagonal, and amounts
to shifting $n \rightarrow n + \alpha$ in the coefficients of
$\Psi_{n}^{\pm(2)}$, which describes both the down component of the
doublet and the singlet. The 4D kinetic Lagrangian for the field
$\Psi^{(i)}=\Psi_+^{(i)} + \Psi_-^{(i)}$, defined as ${\cal L}_{4D} =
\int_0^{2\pi R}{\cal L}_{5D}$, is easily computed and reads
\begin{equation}
{\cal L}_{4D} = \sum_{n=0}^\infty \bar \Psi_{n}^{(1)}
\big[i\dslash_4 - m_{n}^{(1)}] \Psi_{n}^{(1)}
+ \sum_{n=-\infty}^\infty \bar \Psi_{n}^{(2)}
\big[i\dslash_4 - m_{n}^{(2)}] \Psi_{n}^{(2)},
\end{equation}
where
\begin{equation}
m_{n}^{(1)} = \frac nR \;,\;\; m_{n}^{(2)} = \frac{n+\alpha}R \;.
\label{massefun}
\end{equation}

\subsection{Symmetric}

For the symmetric representation, we have ${\cal R}(P)=P\otimes P^T$
and ${\cal R}(T)=T \otimes T^T$ in (\ref{modes}). Doubling again the
modes for convenience, the untwisted wave functions read in this case:
\begin{eqnarray}
\Psi_\pm(y) = \sum_{n=-\infty}^\infty \frac{\eta_n}{\sqrt{2}} \left(
\matrix{
\pm \sqrt{2} f_n^\pm(y)\, \phi_{n}^{\pm a} \a
\pm f_n^\pm(y)\, \phi_{n}^{\pm c} \a
f_n^\mp(y)\, \psi_{n}^{\pm u} \cr
\pm f_n^\pm(y)\, \phi_{n}^{\pm c} \a
\pm \sqrt{2} f_n^\pm(y)\, \phi_{n}^{\pm b} \a
f_n^\mp(y)\, \psi_{n}^{\pm d} \cr
f_n^\mp(y)\, \psi_{n}^{\pm u} \a
f_n^\mp(y)\, \psi_{n}^{\pm d} \a
\pm \sqrt{2}f_n^\pm(y)\, \chi_{n}^{\pm}}
\right)\;,
\label{dousingS}
\end{eqnarray}
where, as before, we denoted the upper and lower components of the
$SU(2)\times U(1)$ doublet by $\psi^u$ and $\psi^d$, the singlet by
$\chi$ and the three components of the triplet by $\phi^a$, $\phi^b$
and $\phi^c$. The diagonal basis is defined by the new fields
\begin{eqnarray}
\Psi_n^{\pm (1)} \a=\a \eta_n \Big(\psi_n^{\pm u}-\phi_n^{\pm c}\Big) \;,\;\;
\Psi_n^{\pm (2)} = \eta_n \Big(\psi_n^{\pm d} +
\frac{\chi_n^{\pm} - \phi_n^{\pm b}}{\sqrt{2}}\Big) \;,\nn \\
\Psi_{n}^{\pm (3)} \a=\a \phi_n^{\pm a} \;,\;\;
\Psi_n^{\pm (4)} = \frac{\chi_n^{\pm} + \phi_n^{\pm b}}{\sqrt{2}} \;,
\label{symmredef}
\end{eqnarray}
where all the modes in $\Psi_n^{\pm (1)}$ and $\Psi_n^{\pm (2)}$ are
now physical. In this way, calling for short $\Psi_n^{\pm (2\pm 4)} =
\Psi_{n}^{\pm (2)} \pm \eta_n \Psi_{n}^{\pm (4)}$, one has
\begin{eqnarray}
\Psi_\pm(y) = \frac 1{\sqrt{2}}\sum_{n=-\infty}^\infty \!\left(\matrix{
\pm \sqrt{2} \eta_n f_n^\pm(y) \Psi_{n}^{\pm (3)} \a
\mp f_n^\pm(y) \Psi_{n}^{\pm (1)} \a
f_n^\mp(y) \Psi_{n}^{\pm (1)} \cr
\mp f_n^\pm(y) \Psi_{n}^{\pm (1)} \a
\mp f_n^\pm(y) \Psi_{n}^{\pm (2-4)} \a
f_n^\mp(y) \Psi_{n}^{\pm (2)} \cr
f_n^\mp(y) \Psi_{n}^{\pm (1)} \a
f_n^\mp(y) \Psi_{n}^{\pm (2)} \a
\pm f_n^\pm(y) \Psi_{n}^{\pm (2+4)}}
\right).
\label{wavesym}
\end{eqnarray}
{}From the above expression, we see that the field $\Psi^{\pm(2)}$ again
appears both in the down component of the doublet and the singlet. The
action of the twist amounts to shifting $n \rightarrow n + \alpha$ in
the coefficients of $\Psi_{n}^{\pm(1)}$ and $n \rightarrow n + 2 \alpha$
in the coefficients of $\Psi_{n}^{\pm(2)}$. The 4D kinetic Lagrangian for
the new fields is
\begin{equation}
{\cal L}_{4D} = \sum_{n=-\infty}^\infty\sum_{i=1,2}
\bar \Psi_{n}^{(i)} \big[i\dslash_4 - m_{n}^{(i)}] \Psi_{n}^{(i)}+
\sum_{n=0}^\infty\sum_{i=3,4} \bar \Psi_{n}^{(i)}
\big[i\dslash_4 - m_{n}^{(i)}]\Psi_{n}^{(i)}\,,
\end{equation}
where
\begin{equation}
m_{n}^{(1)} = \frac {n+\alpha}R \;,\;\;
m_{n}^{(2)} =\frac{n+2\alpha}R \;,\;\;
m_{n}^{(3)} = \frac{n}R \;,\;\;
m_{n}^{(4)} =\frac{n}R \;.
\label{massesym}
\end{equation}

\subsection{Adjoint}

For the adjoint representation, we have ${\cal R}(P)=P\otimes
P^\dagger$ and ${\cal R}(T)=T \otimes T^\dagger$ in (\ref{modes}). The
decomposition of untwisted fields reads
\begin{eqnarray}
\Psi_\pm(y) = \sum_{n=-\infty}^\infty \frac{\eta_n}{\sqrt{2}} \left(
\matrix{
\pm  f_n^\pm(y)\, (Z_{n}^{\pm} \!+\! \frac{1}{\!\sqrt{3}}\chi_n^\pm) \a
\pm f_n^\pm(y)\, Y_{n}^{\pm} \a
f_n^\mp(y)\, \psi_{n}^{\pm u} \cr
\pm f_n^\pm(y)\, (Y_{n}^{\pm })^\dagger \a
\mp f_n^\pm(y)\, (Z_{n}^{\pm} \!-\! \frac{1}{\!\sqrt{3}}\chi_n^\pm) \a
f_n^\mp(y)\, \psi_{n}^{\pm d} \cr
f_n^\mp(y)\, (\psi_{n}^{\pm u})^\dagger \a
f_n^\mp(y)\, (\psi_{n}^{\pm d})^\dagger \a
\pm \frac{2}{\sqrt{3}}f_n^\pm(y)\, \chi_{n}^{\pm}}
\right),
\label{dousingA}
\end{eqnarray}
where, as before, we denoted the upper and lower complex components of
the $SU(2)\times U(1)$ doublet by $\psi^u$ and $\psi^d$, the singlet
by $\chi$ and the three components of the triplet by $Z$ and $Y,Y^\dagger$.
The diagonal basis is defined by
\begin{eqnarray}
\Psi_n^{\pm (1)} \a=\a \eta_n \Big(Y_n^\pm -\psi_n^{\pm u}\Big) \;,\;\;
\Psi_n^{\pm (2)} = \eta_n \Big( {\rm Re}\,\psi_n^{\pm d} +
\frac{Z_n^{\pm} - \sqrt{3}\chi_n^{\pm}}{2}\Big) \;,\nn \\
\Psi_n^{\pm (3)} \a= \a \frac{\sqrt{3}Z_n^{\pm} + \chi_n^{\pm}}{2} \;,\;\;
\Psi_{n}^{\pm (4)} = {\rm Im}\, \psi_n^{\pm d} \;,
\label{adjredef}
\end{eqnarray}
where all modes of $\Psi_n^{\pm (1)}$ and $\Psi_n^{\pm (2)}$ are
physical and the first is a complex field. Defining for short
$\Psi_n^{\pm (2\pm 3)} = \Psi_n^{\pm(2)} \pm \frac{\eta_n}{\sqrt{3}}
\Psi_n^{\pm(3)}$ and $\Psi_n^{\pm (2\pm 4)} = \Psi_n^{\pm(2)} \pm i
\eta_n \Psi_n^{\pm(4)}$, we obtain, in the new basis:
\begin{eqnarray}
\Psi_\pm(y) = \frac 1{\sqrt{2}}\sum_{n=-\infty}^\infty \!\left(\matrix{
\pm \frac{2}{\sqrt{3}} \eta_n f_n^\pm(y) \Psi_{n}^{\pm (3)} \a
\pm f_n^\pm(y) \Psi_{n}^{\pm (1)} \a
f_n^\mp(y) \Psi_{n}^{\pm (1)} \cr
\pm f_n^\pm(y) (\Psi_{n}^{\pm (1)})^\dagger \a
\mp f_n^\pm(y) \Psi_{n}^{\pm (2+3)} \a
f_n^\mp(y) \Psi_{n}^{\pm(2+4)} \cr
f_n^\mp(y) (\Psi_{n}^{\pm (1)})^\dagger \a
f_n^\mp(y) (\Psi_{n}^{\pm(2+4})^\dagger \a
\mp f_n^\pm(y) \Psi_{n}^{\pm (2-3)}}
\right) \;.
\label{waveadj}
\end{eqnarray}
The action of the twist amounts to shifting $n \rightarrow n + \alpha$
in the coefficients of $\Psi_{n}^{\pm(1)}$ and $n \rightarrow n + 2 \alpha$
in the coefficients of $\Psi_{n}^{\pm(2)}$. The 4D Lagrangian for the new
fields is
\begin{eqnarray}
{\cal L}_{4D} = \a\a \sum_{n=-\infty}^\infty
(\bar \Psi_{n}^{(1)})^\dagger \big[i\dslash_4 - m_{n}^{(1)}] \Psi_{n}^{(1)}
+ \sum_{n=-\infty}^\infty
\bar \Psi_{n}^{(2)} \big[i\dslash_4 - m_{n}^{(2)}] \Psi_{n}^{(2)}\nn \\
\a\a + \sum_{n=0}^\infty\sum_{i=3,4}
\bar \Psi_{n}^{(i)} \big[i\dslash_4 - m_{n}^{(i)}] \Psi_{n}^{(i)}
\end{eqnarray}
where
\begin{equation}
m_{n}^{(1)} = \frac {n+\alpha}R \;,\;\;
m_{n}^{(2)} =\frac{n+2\alpha}R \;,\;\;
m_{n}^{(3)} = \frac{n}R \;,\;\;
m_{n}^{(4)} =\frac{n}R \;.
\label{masseadj}
\end{equation}


\begin{thebibliography}{99}

\small

\bibitem{Antoniadis}
I.~Antoniadis,
Phys.\ Lett.\ B {\bf 246} (1990) 377; \\
I.~Antoniadis, C.~Munoz, M.~Quiros,
Nucl.\ Phys.\ B {\bf 397} (1993) 515
[hep-ph/9211309]; \\
A.~Delgado, A.~Pomarol, M.~Quiros,
Phys.\ Rev.\ D {\bf 60} (1999) 095008
[hep-ph/9812489];\\
I.~Antoniadis, S.~Dimopoulos, A.~Pomarol, M.~Quiros,
Nucl.\ Phys.\ B {\bf 544} (1999) 503
[hep-ph/9810410].

\bibitem{early}
N.~S.~Manton,
Nucl.\ Phys.\ B {\bf 158} (1979) 141;\\
D.~B.~Fairlie,
Phys.\ Lett.\ B {\bf 82} (1979) 97;
J.\ Phys.\ G {\bf 5} (1979) L55;\\
P.~Forgacs, N.~S.~Manton,
Commun.\ Math.\ Phys.\  {\bf 72} (1980) 15;\\
S.~Randjbar-Daemi, A.~Salam, J.~Strathdee,
Nucl.\ Phys.\ B {\bf 214} (1983) 491;\\
N.~V.~Krasnikov,
Phys.\ Lett.\ B {\bf 273} (1991) 246.

\bibitem{Hatanaka:1998yp}
H.~Hatanaka, T.~Inami, C.~Lim,
Mod.\ Phys.\ Lett.\ A {\bf 13} (1998) 2601
[hep-th/9805067].

\bibitem{Dvali:2001qr}
G.~R.~Dvali, S.~Randjbar-Daemi, R.~Tabbash,
Phys.\ Rev.\ D {\bf 65} (2002) 064021
[hep-ph/0102307].

\bibitem{Hall:2001zb}
L.~J.~Hall, Y.~Nomura, D.~R.~Smith,
Nucl.\ Phys.\ B {\bf 639} (2002) 307
[hep-ph/0107331].

\bibitem{Antoniadis:2001cv}
I.~Antoniadis, K.~Benakli, M.~Quiros,
New J.\ Phys.\  {\bf 3} (2001) 20
[hep-th/0108005].

\bibitem{Csaki:2002ur}
C.~Csaki, C.~Grojean, H.~Murayama,
[hep-ph/0210133].

\bibitem{Burdman:2002se}
G.~Burdman, Y.~Nomura,
[hep-ph/0210257].

\bibitem{Dixon}
L.~J.~Dixon, J.~A.~Harvey, C.~Vafa, E.~Witten,
Nucl.\ Phys.\ B {\bf 261} (1985) 678.

\bibitem{vonGersdorff}
G.~von Gersdorff, N.~Irges, M.~Quiros,
[hep-ph/0206029];
Nucl.\ Phys.\ B {\bf 635} (2002) 127
[hep-th/0204223];
Phys.\ Lett.\ B {\bf 551} (2003) 351
[hep-ph/0210134].

\bibitem{Wilson-line}
E.~Witten,
Nucl.\ Phys.\ B {\bf 258} (1985) 75; \\
L.~E.~Ib\'a\~nez, H.~P.~Nilles, F.~Quevedo,
Phys.\ Lett.\ B {\bf 187} (1987) 25.

\bibitem{hos}
Y. Hosotani,
Phys.\ Lett.\ B {\bf 126} (1983) 309;
Phys.\ Lett.\ B {\bf 129} (1983) 193;
Ann.\ Phys.\ {\bf 190} (1989) 233.

\bibitem{SS}
J.~Scherk, J.~H.~Schwarz,
Phys.\ Lett.\ B {\bf 82} (1979) 60;
Nucl.\ Phys.\ B {\bf 153} (1979) 61.

\bibitem{NDA1}
S.~Weinberg,
Physica A {\bf 96} (1979) 327;\\
A.~Manohar and H.~Georgi,
Nucl.\ Phys.\ B {\bf 234} (1984) 189.

\bibitem{NDA2}
Z.~Chacko, M.~A.~Luty and E.~Ponton,
JHEP {\bf 0007} (2000) 036
[hep-ph/9909248].

\bibitem{Kubo:2001zc}
M.~Kubo, C.~S.~Lim, H.~Yamashita,
[hep-ph/0111327].

\bibitem{Haba:2002py}
N.~Haba, M.~Harada, Y.~Hosotani, Y.~Kawamura,
[hep-ph/0212035].

\bibitem{Delgado:1999sv}
A.~Delgado, A.~Pomarol, M.~Quiros,
JHEP {\bf 0001} (2000) 030
[hep-ph/9911252].

\bibitem{Carena:2002me}
M.~Carena, T.~M.~Tait, C.~E.~Wagner,
Acta Phys.\ Polon.\ B {\bf 33} (2002) 2355
[hep-ph/0207056].

\bibitem{Antoniadis:1993jp}
I.~Antoniadis and K.~Benakli,
Phys.\ Lett.\ B {\bf 326} (1994) 69
[arXiv:hep-th/9310151].

\bibitem{consistency}
C.~Kounnas, M.~Porrati,
Nucl.\ Phys.\ B {\bf 310} (1988) 355; \\
S.~Ferrara, C.~Kounnas, M.~Porrati, F.~Zwirner,
Nucl.\ Phys.\ B {\bf 318} (1989) 75.

\bibitem{HH}
L.~J.~Hall, Y.~Nomura,
Phys.\ Rev.\ D {\bf 64} (2001) 055003
[hep-ph/0103125]; \\
A.~Hebecker, J.~March-Russell,
Nucl.\ Phys.\ B {\bf 613} (2001) 3
[hep-ph/0106166].

\bibitem{ms3}
A.~Masiero, C.~A.~Scrucca, M.~Serone, L.~Silvestrini,
Phys.\ Rev.\ Lett.\  {\bf 87} (2001) 251601
[hep-ph/0107201].

\bibitem{GS}
M.~B.~Green, J.~H.~Schwarz,
Phys.\ Lett.\ B {\bf 149} (1984) 117; \\
M.~Dine, N.~Seiberg, E.~Witten,
Nucl.\ Phys.\ B {\bf 289} (1987) 589.

\bibitem{anomaly}
N.~Arkani-Hamed, A.~G.~Cohen, H.~Georgi,
Phys.\ Lett.\ B {\bf 516} (2001) 395
[hep-th/0103135]; \\
C.~A.~Scrucca, M.~Serone, L.~Silvestrini, F.~Zwirner,
Phys.\ Lett.\ B {\bf 525} (2002) 169
[hep-th/0110073].

\bibitem{CS}
R.~Barbieri, R.~Contino, P.~Creminelli, R.~Rattazzi, C.~A.~Scrucca,
Phys.\ Rev.\ D {\bf 66} (2002) 024025
[hep-th/0203039]; \\
L.~Pilo, A.~Riotto,
Phys.\ Lett.\ B {\bf 546} (2002) 135
[hep-th/0202144].

\bibitem{CSnonab}
H.~D.~Kim, J.~E.~Kim, H.~M.~Lee,
JHEP {\bf 0206} (2002) 048 [hep-th/0204132].

\bibitem{sst}
C.~A.~Scrucca, M.~Serone, M.~Trapletti,
Nucl.\ Phys.\ B {\bf 635} (2002) 33
[hep-th/0203190].

\bibitem{slansky}
R.~Slanski,
Phys.\ Rep.\ B {\bf 79} (1981) 1.

\bibitem{Dvali:2000rx}
G.~R.~Dvali, G.~Gabadadze, M.~A.~Shifman,
Phys.\ Lett.\ B {\bf 497} (2001) 271
[hep-th/0010071].

\bibitem{delAguila:2003bh}
F.~del Aguila, M.~Perez-Victoria, J.~Santiago,
[hep-th/0302023].

\bibitem{abd}
I.~Antoniadis, C.~Bachas, E.~Dudas,
Nucl.\ Phys.\ B {\bf 560} (1999) 93
[hep-th/9906039].

\bibitem{Ponton:2001hq}
E.~Ponton and E.~Poppitz,
JHEP {\bf 0106} (2001) 019
[hep-ph/0105021].

\end{thebibliography}
\end{document}